\documentclass[twocolumn, twocolappendix]{aastex631}

\usepackage{stix, isomath, booktabs, longtable, enumitem, amsmath, multirow, threeparttable, float}
\let\tablenum\relax
\usepackage{siunitx}
\usepackage[version=4, arrows=pgf-filled]{mhchem}
\usepackage{anyfontsize}

\newcommand{\RJ}{\textit{R}_J}

\newcommand{\OI}{O\,\textsc{i}}
\newcommand{\OII}{O\,\textsc{ii}}
\renewcommand{\SI}{S\,\textsc{i}}
\newcommand{\SII}{S\,\textsc{ii}}
\newcommand{\NaI}{Na\,\textsc{i}}
\newcommand{\KI}{K\,\textsc{i}}

\received{January 27, 2025}
\revised{May 19, 2025}
\accepted{June 1, 2025}

\begin{document}

\title{Detection of New Auroral Emissions at Io and Implications for Its Interaction with the Plasma Torus}

\correspondingauthor{Zachariah Milby}
\email{zmilby@caltech.edu}

\author[0000-0001-5683-0095]{Zachariah {Milby}}
\affiliation{Division of Geological and Planetary Sciences,
California Institute of Technology, USA}

\author[0000-0002-9068-3428]{Katherine {de~Kleer}}
\affiliation{Division of Geological and Planetary Sciences,
California Institute of Technology, USA}

\author[0000-0002-6917-3458]{Carl {Schmidt}}
\affiliation{Center for Space Physics, Boston University, USA}

\begin{abstract}
We observed Io's optical aurora in eclipse on six nights between 2022 and 2024 using Keck I/HIRES. Spectra revealed 13 new auroral emissions not identified previously, tripling the total number of optical emissions lines detected at Io. These included the \OI{} lines at 777.4 and 844.6~nm, the \NaI{} lines at 818.3 and 819.5~nm, the [\SI{}] lines at 458.9 and 772.5~nm, the \SI{} triplet at 922.3~nm, the [\OII{}] lines at 732.0 and 733.0~nm and the [\SII{}] lines at 406.9, 407.6, 671.6 and 673.1~nm. We leveraged these new detections by comparing with imaging data from the 2001 Cassini flyby to better understand the distribution of atmospheric species and their contribution to the observed auroral brightnesses. Our auroral emission model showed that the observed 557.7, 777.4 and 844.6~nm oxygen emission line brightnesses could be explained by excitation by electron impact of canonical 5~eV torus electrons on an atmosphere composed of \ce{O}, \ce{SO2} and an isoelectronic proxy for \ce{SO}. The \ce{SO2} emission did not decrease immediately after eclipse ingress, suggesting the emitting column may be restricted to higher altitudes. The derived $\ce{O}/\ce{SO2}$ mixing ratio was typically about 10\%, but it also exhibited order-of-magnitude variance during some observations. Io's 630.0~nm [O\,\textsc{i}] brightness did not strongly vary with plasma sheet distance, suggesting electron flux at Io varies substantially beyond model predictions.
\end{abstract}


\section{Introduction}

A combination of volcanic outgassing and sublimation of surface frost by sunlight produces Io's thin atmosphere. Although sulfur dioxide (\ce{SO2}) constitutes the bulk of Io's atmosphere \citep[first detected by][]{Pearl1979}, observations have revealed the presence of smaller concentrations of sulfur monoxide \citep[\ce{SO},][]{Lellouch1996}, atomic sulfur (\ce{S}) and atomic oxygen \citep[\ce{O},][]{Ballester1987} in addition to the alkali compounds sodium chloride \citep[\ce{NaCl},][]{Lellouch2003} and potassium chloride \citep[\ce{KCl},][]{Moullet2013} and their dissociation products atomic sodium \citep[\ce{Na},][]{Schneider1987}, atomic potassium \citep[\ce{K},][]{Brown2001} and atomic chlorine \citep[\ce{Cl},][]{Feaga2004}. \citet{Spencer2000} detected \ce{S2} on one occasion in a volcanic plume, but it has not yet been observed elsewhere in Io's atmosphere. Observational results interpreted with photochemical models \citep[e.g.,][]{Summers1996,Feldman2000,Moses2002,Dols2024} present Io's atmosphere as spatially variable in both density and composition, with global coronae made of \ce{S} and \ce{O} along with a higher density molecular \ce{SO2} and \ce{SO} atmosphere concentrated near equatorial latitudes below $\pm\ang{30}$ with a column density 30 to 60 times larger than the \ce{SO2} corona \citep{Strobel2001}. It remains uncertain what fraction of SO is nascent volcanic outgassing \citep{deKleer2019} versus fragments of \ce{SO2} dissociation \citep{dePater2020}, though models of Io's atmosphere \citep[e.g.,][]{Geissler2004,Roth2011,Dols2024} typically assume a 10\% mixing ratio with \ce{SO2} \citep{McGrath2000}, consistent with observed ratios between 3 and 10\% \citep{Lellouch1996,Moullet2010}.

Solar insolation produces collisional densities within the low-latitude equatorial atmosphere on the dayside, while on the nightside (and during eclipse) \ce{SO2} can freeze back onto the surface, resulting in a thin exosphere \citep{Saur2004,Tsang2016,dePater2023}. The differing spatial distributions between Io's molecular and atomic atmospheres help to explain spatially-resolved observations of Io's aurora in eclipse which show three distinct morphological features: (1) a diffuse limb glow, (2) an extended coronal emission and (3) bright, isolated emissions near the equator \citep[e.g.,][]{Geissler1999,Geissler2004,Roesler1999,Retherford2000,Retherford2003,Retherford2007,Roth2011,Roth2014,Saur2000}. Observations with fine spatial resolution also show emission from volcanic features like plumes and vents \citep{Geissler1999,Geissler2004}. The equatorial spots shift in latitude as Jupiter rotates, following tangents between Jupiter's magnetic field and Io's surface, with a small modification to the local field due to electromagnetic induction from within Io's core \citep{Roesler1999,Roth2014,Roth2017}. The limb glow has been most readily identified at the poles, where the observed north-south brightness asymmetry favors the hemisphere facing the centrifugal equator of the Io plasma torus \citep{Retherford2003,Moore2010}.

These different auroral morphologies are probably due to variation in both the spatial distribution and composition of Io's atmosphere as well as the properties of the excitation mechanisms for the the specific emissions. However, no studies have yet quantitatively determined the relative contributions of atomic and/or molecular species responsible for the auroral emissions at optical wavelengths. Though the observed emissions are primarily from atoms and atomic ions (O, S, Cl, Na and K), dissociation of molecules and molecular ions can also produce excited fragments. For instance, at Europa and Ganymede, most of the optical atomic oxygen auroral emission in eclipse comes from dissociative electron impact on \ce{O2}; electron impact on \ce{O} is only a minor contribution \citep{deKleer2018,deKleer2019,deKleer2023,Milby2024}. \citet{Bouchez2000} found that the impact of 5~eV electrons on an atomic oxygen column of approximately $10^{15}~\mathrm{cm^{-2}}$ could explain the auroral brightness ratios they observed at Io, though they note that \citet{Scherb1993} suggested that dissociative excitation of oxygen from electron impact on \ce{SO} and \ce{SO2} might constitute the majority of the forbidden oxygen emissions, an interpretation shared by \citet{Oliversen2001}. \citet{Geissler2004} concluded that most of the emissions captured in their near-UV/blue broadband filter images likely came from electron-impact fluorescence of \ce{SO2}. \citet{Geissler2001} came to a similar conclusion earlier by observing emission enhancement near active volcanic vents in broadband clear filter images. \citet{Schmidt2023} concluded that the $557.7/630.0$~nm [\OI{}] ratio could be explained by electron impact on atomic oxygen, but they also noted the ratio required less energetic electrons than the 5~eV ambient population upstream of Io in the warm torus.

High-cadence observations leverage the large collecting areas of ground-based telescopes like the twin Keck telescopes on the summit of Maunakea to achieve high signal-to-noise in short integration times, but atmospheric seeing limits their spatial resolution. In contrast, space-based observatories like the Hubble Space Telescope (HST) can achieve better spatial resolution at UV wavelengths, but its short orbital period and small mirror area limits the signal-to-noise achievable over the limited-duration satellite eclipses. Spacecraft flyby imaging of Io in eclipse by Galileo \citep{Geissler1999,Geissler2001} and Cassini \citep{Geissler2004} produced images with good spatial resolution, but the broad filter bandpasses made it difficult to identify individual emissions and their relative contributions. Nevertheless these images allowed for the characterization of distinct emission morphologies. \citet{Bouchez2000} reported the first eclipse detections of individual components of Io's optical aurora, including the 557.7, 630.0 and 636.4~nm [\OI{}] lines and the sodium D lines at 589.0 (D$_1$) and 589.6~nm (D$_2$). \citet{Schmidt2023} reported the first eclipse detections of the potassium D lines at 766.5~nm (D$_2$) and 769.9~nm (D$_1$) and showed that the sodium D line brightness and line width exhibits a systematic temporal response to Io's passage through Jupiter's shadow.

We used high-resolution optical spectra of Io's auroral emission in eclipse as a remote sensing window into the interaction between Io's atmosphere and electrons within Jupiter's magnetosphere. We compared both newly and previously identified atomic emission lines to broadband filter images of Io in eclipse taken during the Cassini flyby of the Jovian system on 2001 January 05 to determine which species contributed to which discrete aurora features and whether the atomic emissions identified in our spectra could reasonably account for all the brightness observed in the broadband images. We used an auroral emission model to evaluate the observed brightnesses under the assumption of electron-impact excitation to determine both the atmospheric species contributing to the observed atomic emissions (including both direct atomic excitation and electron-impact dissociation producing excited atomic species) and the energy of the electrons exciting the aurora. We evaluated the connection between Io's brightest oxygen aurora and Io's physical position within the plasma sheet to determine if the primary driver of the absolute magnitude of Io's aurora is the ambient density of the electrons. Finally, we evaluated whether certain ion emission lines could be used to probe electron density.

\section{Observations and Data Reduction}

\begin{figure*}
    \centering
    \includegraphics[width=\textwidth]{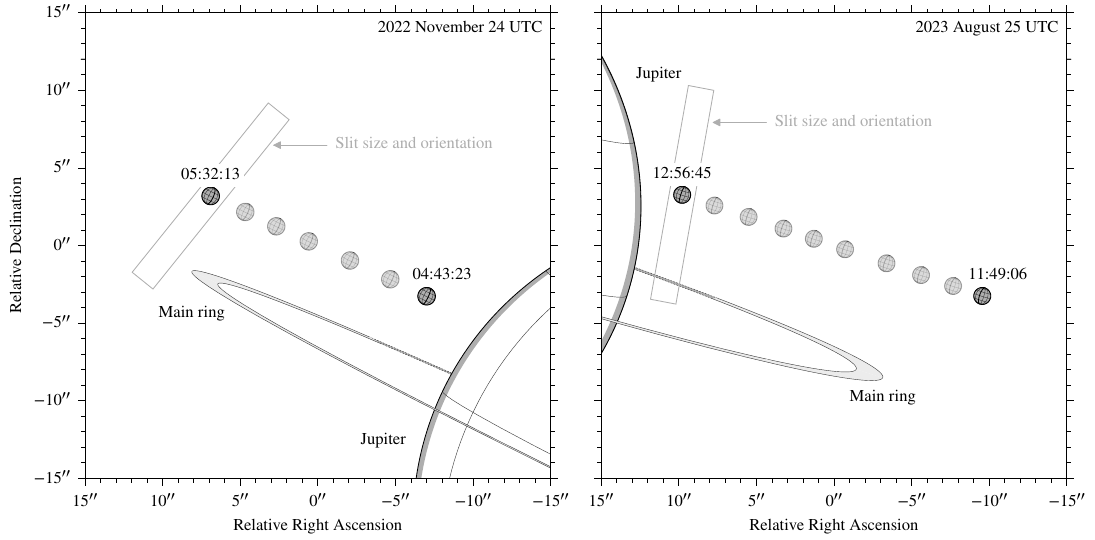}
    \caption{Example viewing geometry of the apparent motion of Io in eclipse relative to Jupiter for egress observations on 2022 November 24 (left) and ingress observations on 2023 August 25 (right). Each drawing of Io shows its position relative to Jupiter for each of the spectra taken on these nights, and the annotated times show the UTC time at the start of the first and last observations. The last observation also includes the angular projection of the slit on the sky and its orientation (the slit size and orientation were fixed for all observations on a given night).}
    \label{fig:viewing-geometry}
\end{figure*}

We analyzed 40 spectra of Io in eclipse by Jupiter taken over six nights between 2022 and 2024. These spectra were acquired with the High Resolution Echelle Spectrometer \citep[HIRES;][]{Vogt1994}, an optical wavelength echelle spectrograph mounted on the Nasmyth platform of the Keck I telescope on the summit of Maunakea. HIRES has two cross dispersers (called HIRESb and HIRESr) which permit observations at either end of the visible spectrum. The spectra acquired on 2023 August 09 UTC used the HIRESb cross disperser, which was optimized for observations in the second order from about 300 to 700~nm. The data acquired on the other five nights used the HIRESr cross disperser, which was optimized for observations in the first order from about 400 to 1000~nm. For all six nights, we used a slit with a projected angular size on the sky of $\ang{;;1.722}\times\ang{;;14}$.

We used the same observing methodology employed by previous HIRES optical aurora studies \citep{Bouchez2000,deKleer2018,deKleer2023,Schmidt2023,Milby2024}. During the 2022 observations, Jupiter was near eastern quadrature relative to Earth and the telescope line of sight captured Io as it emerged from behind Jupiter's disk through eclipse egress. We used Ganymede as the guide satellite for these observations. During the 2023 and 2024 observations, Jupiter was near western quadrature and the line of sight captured Io during eclipse ingress until it eventually disappeared behind Jupiter's apparent disk. On all of these nights we used Europa as the guide satellite. Figure \ref{fig:viewing-geometry} shows an example of the viewing geometry for both eclipse ingress and egress observations.

We reduced and calibrated the data using the latest versions of the data reduction and calibration pipelines described by \cite{Milby2024}. When Io appears near to Jupiter's limb, scattered solar continuum from Jupiter's atmosphere can contribute a large background flux with variable Doppler shift along the slit. This makes the background very difficult to characterize and remove, especially at near-infrared wavelengths greater than 875~nm, where CCD fringing complicates the background structure. As a result, we were only able to retrieve the brightness of the 922.3~nm \SI{} triplet from the 2022 November 24 data and we removed two spectra taken on 2023 August 25 (timestamps 2023-08-25T11:49:06.000 and 2023-08-25T12:56:45.000) and one spectrum taken on 2024 October 21 (timestamp 2024-10-21T12:04:28.000) from our analysis.

Keck I pointing and tracking becomes increasingly unreliable at high elevations near the zenith. Because we used manual offsets and tracking rates in order to integrate on the otherwise invisible eclipsed satellite, during some observations taken at high elevation the eclipsed satellite may move along the spatial axis of the slit or slip out of it all together. We identified this problem in three spectra taken on 2024 October 05 (timestamps 2024-10-05T14:02:13.000, 2024-10-05T14:23:11.000 and 2024-10-05T14:30:14.000) and subsequently removed them from our analysis.

In addition to the science exposures of Io in eclipse and the fully-illuminated guide satellite, we took sets of CCD bias exposures, flat lamp exposures, thorium-argon (ThAr) arc lamp exposures and a trace spectrum of a standard star for data reduction and wavelength calibration purposes. For calibration from detector counts to physical units, we used a spectrum of Jupiter's meridian following the same methodology as \cite{deKleer2018,deKleer2019,deKleer2023} and \citet{Milby2024}. These data are available from the Keck Observatory Archive (KOA)\footnote{\url{https://koa.ipac.caltech.edu/}} and the full list of data files used in this study are listed in Tables \ref{tab:files-2022-11-24} through \ref{tab:files-2024-10-21} in the Appendix.

\begin{table*}
\centering
\caption{Overview of the Keck/HIRES Observations of Io in Eclipse}
\label{tab:observation-information}
\begin{tabular}{llcS[table-format=-1.3, table-text-alignment=center]@{\hspace{1ex}}c@{\hspace{1ex}}S[table-format=-1.3, table-text-alignment=center]S[table-format=3.1, table-text-alignment=center]@{\hspace{1ex}}c@{\hspace{1ex}}S[table-format=3.1, table-text-alignment=center]cS[table-format=-1.3, table-text-alignment=center]@{\hspace{1ex}}c@{\hspace{1ex}}S[table-format=-1.3, table-text-alignment=center]S[table-format=-2.1, table-text-alignment=center]@{\hspace{1ex}}c@{\hspace{1ex}}S[table-format=-2.1, table-text-alignment=center]}
\toprule
{Date\textsuperscript{\scriptsize{a}}} \relax& \multirow{2}{*}{{Cross Disperser\textsuperscript{\scriptsize{b}}}} \relax&  \multirow{2}{*}{Used / Total\textsuperscript{\scriptsize{c}}} \relax&
\multicolumn{3}{c}{$\lambda_\mathrm{m}$\textsuperscript{\scriptsize{d}}} \relax& \multicolumn{3}{c}{$\phi_\mathrm{m}$\textsuperscript{\scriptsize{e}}} \relax& {$r_\mathrm{I}$\textsuperscript{\scriptsize{f}}} \relax& \multicolumn{3}{c}{$d$\textsuperscript{\scriptsize{g}}} \relax& \multicolumn{3}{c}{$v_\mathrm{rel}$\textsuperscript{\scriptsize{h}}}\\
{[UTC]} \relax&  \relax&  \relax& \multicolumn{3}{c}{[deg]} \relax& \multicolumn{3}{c}{[deg]} \relax& {[\unit{\RJ}]} \relax& \multicolumn{3}{c}{[\unit{\RJ}]} \relax& \multicolumn{3}{c}{[\unit{km.s^{-1}}]}\\
\midrule
2022 November 24 \relax& HIRESr \relax& 7 / 7 \relax& -8.929 \relax& to \relax& -9.526 \relax& 359.6 \relax& to \relax& 337.0 \relax& 5.877 \relax& -0.633 \relax& to \relax& -0.676 \relax& 21.0 \relax& to \relax& 19.1\\
2023 August 09 \relax& HIRESb \relax& 7 / 7 \relax& 3.230 \relax& to \relax& 0.043 \relax& 88.9 \relax& to \relax& 69.2 \relax& 5.913 \relax& 0.232 \relax& to \relax& 0.003 \relax& -21.3 \relax& to \relax& -23.0\\
2023 August 25 \relax& HIRESr \relax& 8 / 10 \relax& -3.628 \relax& to \relax& 1.479 \relax& 271.5 \relax& to \relax& 240.1 \relax& 5.910 \relax& -0.261 \relax& to \relax& 0.106 \relax& -19.7 \relax& to \relax& -22.4\\
2024 September 12 \relax& HIRESr \relax& 7 / 7 \relax& -5.936 \relax& to \relax& -8.259 \relax& 30.0 \relax& to \relax& 8.1 \relax& 5.921 \relax& -0.430 \relax& to \relax& -0.596 \relax& -21.4 \relax& to \relax& -23.3\\
2024 October 05 \relax& HIRESr \relax& 4 / 7 \relax& 9.405 \relax& to \relax& 9.390 \relax& 168.6 \relax& to \relax& 148.8 \relax& 5.917 \relax& 0.671 \relax& to \relax& 0.670 \relax& -19.1 \relax& to \relax& -20.7\\
2024 October 21 \relax& HIRESr \relax& 4 / 5 \relax& -9.441 \relax& to \relax& -9.093 \relax& 342.8 \relax& to \relax& 323.0 \relax& 5.913 \relax& -0.679 \relax& to \relax& -0.654 \relax& -15.9 \relax& to \relax& -17.6\\
\bottomrule
\multicolumn{16}{p{0.92\linewidth}}{\textbf{Notes.} Inaccuracies in telescope tracking at high elevations caused Io to move along and/or out of the slit during three observations on 2024 October 05 UTC, so we removed them from our analysis. These observations have UTC timestamps of 2024-10-05T14:02:13.000, 2024-10-05T14:23:11.000 and 2024-10-05T14:30:14.000. Strong Jovian scattered light background resulted in poor background subtraction for the first and last spectra taken on 2023 August 25 UTC and the first spectrum taken on 2024 October 21, so we removed those observations (2023-08-25T11:49:06.000, 2023-08-25T12:56:45.000 and 2024-10-21T12:04:28.000) as well.}\\[-5pt]
\multicolumn{16}{p{0.92\linewidth}}{
\begin{itemize}[nosep, labelsep=-1.5pt, align=left, leftmargin=*]
\item[\textsuperscript{a}]UTC date on Maunakea at the start of the observations.
\item[\textsuperscript{b}]HIRES cross disperser used.
\item[\textsuperscript{c}]Number of spectra used out of total number of spectra taken.
\item[\textsuperscript{d}]Magnetospheric latitudes of Io over the range of observations.
\item[\textsuperscript{e}]Magnetospheric longitude of Io over the range of observations. This is the same as the System\,\textsc{iii} west longitude, but converted here to east longitude.
\item[\textsuperscript{f}]Io's orbital distance from Jupiter.
\item[\textsuperscript{g}]Distance between Io and the plasma sheet centrifugal equator; positive when Io is above the midplane and negative when Io is below the midplane.
\item[\textsuperscript{h}]Velocity of Io relative to an observer on Earth (the negative sign indicates motion toward the observer).
\end{itemize}
}
\end{tabular}

\end{table*}

\subsection{Systematic Uncertainties}
\citet{Milby2024} analyzed similar HIRES spectra of Ganymede in eclipse and found the $630.0~\mathrm{nm}/636.4~\mathrm{nm}$ [\OI{}] emission ratio deviated from the optically-thin ratio in a way that could not be explained by collisional deexcitation. As a result, they added a 9\% systematic uncertainty to all of their auroral brightness measurements. We evaluated the same $630.0~\mathrm{nm}/636.4~\mathrm{nm}$ [\OI{}] emission ratio in the Io eclipse spectra to determine if we needed to include any systematic error. However, we found the ratio was within one standard deviation of the expected ratio of 3.09 \citep{Wiese1996} and subsequently did not include any additional systematic uncertainty for any of the auroral brightnesses reported in this study.

\section{Analysis}

\subsection{Average Brightnesses}
All of the brightnesses listed in this study are disk-integrated and then averaged across all spectra used for each night. Though each measurement is independent, there is inherent variability in the brightnesses beyond expected Poisson noise due to short-timescale changes in the number density of the electrons exciting the auroral emissions and/or the atmospheric column densities. We therefore chose to calculate weighted brightness averages rather than arithmetic brightness averages so that measurements with relatively larger uncertainties are appropriately downweighted. We calculated the weighed average brightness $\bar{B}$ of $n$ brightness measurements as
\begin{equation}
    \bar{B} = \frac{\sum_{i=1}^n B_i w_i}{\sum_{i=1}^n w_i},
\end{equation}
where $B_i$ is the $i$th measurement and $w_i = 1/\sigma_i^2$ is the inverse of the square of its corresponding measured uncertainty $\sigma_i$ (equivalently the inverse variance of $B_i$). The propagated uncertainty we report for the weighted average $\bar{\sigma}$ is
\begin{equation}
    \bar{\sigma} = \frac{1}{\sqrt{\sum_{i=1}^n w_i}}.
\end{equation}

\subsection{Auroral Emission Model}\label{sec:aurora-emission-model}
We have further expanded the auroral emission model used by \citet{deKleer2023} and \citet{Milby2024} with additional cross sections relevant to Io's atmospheric composition. For electron impact on \ce{SO2} we have included emission cross sections at 130.4 and 135.6~nm \citep{VattiPalle2004}. Because the 777.4 and 844.6~nm emissions cascade into these UV lines, \citet{Ajello2008} indicated the 777.4 and 844.6~nm emission cross sections could be approximated by scaling the 130.4 and 135.6~nm cross sections using values they provided. We used the excitation cross section for electron impact on \ce{SO2} producing \ce{O(^1S)} from \citet{Kedzierski2000} to calculate emission cross sections for 297.2 and 557.7~nm using their relative emission probabilities \citep{Wiese1996}. To date, no measurements of excitation cross sections for electron impact on \ce{SO2} producing \ce{O(^1D)} or similar emission cross sections for [\OI{}] 630.0 or 636.4~nm have been published, however, the \citet{Kedzierski2000} 557.7~nm cross-section quantifies the cascade contribution, effectively setting a lower limit. No emission or excitation cross sections have been published for electron impact on \ce{SO} \citep{McConkey2008}. 

We also used ChantiPy, the Python interface to the CHIANTI atomic database v10.1 \citep{Dere1997,Dere2023}, to calculate photon emission ratios from atomic and ionic columns as a function of electron energy and density. CHIANTI includes many of the emissions listed in Table \ref{tab:all-lines}, with the exception of those from the neutral alkali atoms \NaI{} and \KI{} and the electric dipole \OI{} transitions at 777.4 and 844.6~nm and \SI{} transition at 922.3~nm.

\subsection{Auroral Emission Line Detections}

\begin{table*}
\centering
\caption{Average Auroral Emission Line Brightnesses}
\label{tab:all-lines}
\begin{tabular}{llcccccc}
\toprule
 \relax& \relax& \multicolumn{6}{c}{Disk-Integrated Brightness} \\
\cmidrule(lr){3-8}
Wavelength & \multirow{2}{*}{Species} &  2022 Nov.~24 & 2023 Aug.~09 & 2023 Aug.~25 & 2024 Sep.~12 & 2024 Oct.~05 & 2024 Oct.~21 \\
\multicolumn{1}{c}{[nm]} \relax& & [R] & [R] & [R] & [R] & [R] & [R] \\
\midrule388.4 \relax& \protect[Na~\textsc{i}] \relax& \relax& $< 10$ \relax& \relax& \relax& \relax&  \\
406.9 \relax& \protect[S~\textsc{ii}] \relax& \relax& $310 \pm 18$ \relax& \relax& \relax& \relax&  \\
407.6 \relax& \protect[S~\textsc{ii}] \relax& \relax& $87 \pm 11$ \relax& \relax& \relax& \relax&  \\
458.9 \relax& \protect[S~\textsc{i}] \relax& \relax& $100 \pm 8$ \relax& \relax& \relax& \relax&  \\
464.2 \relax& \protect[K~\textsc{i}] \relax& \relax& $< 15$ \relax& \relax& \relax& \relax&  \\
557.7\textsuperscript{\scriptsize{a}} \relax& \protect[O~\textsc{i}] \relax& $395 \pm 7$ \relax& $461 \pm 12$ \relax& $417 \pm 7$ \relax& $481 \pm 10$ \relax& $440 \pm 12$ \relax& $444 \pm 12$  \\
589.0\textsuperscript{\scriptsize{a}} \relax& Na~\textsc{i} \relax& $16200 \pm 300$ \relax& $12100 \pm 300$ \relax& $18820 \pm 150$ \relax& $20000 \pm 200$ \relax& $14220 \pm 160$ \relax& $14610 \pm 180$  \\
589.6\textsuperscript{\scriptsize{a}} \relax& Na~\textsc{i} \relax& $6340 \pm 180$ \relax& $5070 \pm 140$ \relax& $8810 \pm 90$ \relax& $9160 \pm 110$ \relax& $7500 \pm 90$ \relax& $7820 \pm 110$  \\
630.0\textsuperscript{\scriptsize{a}} \relax& \protect[O~\textsc{i}] \relax& $5430 \pm 30$ \relax& $6090 \pm 110$ \relax& $5630 \pm 40$ \relax& $6700 \pm 50$ \relax& $6130 \pm 50$ \relax& $6160 \pm 50$  \\
636.4\textsuperscript{\scriptsize{a}} \relax& \protect[O~\textsc{i}] \relax& $1768 \pm 15$ \relax& $2200 \pm 20$ \relax& $1903 \pm 12$ \relax& $2361 \pm 19$ \relax& $2092 \pm 18$ \relax& $1950 \pm 20$  \\
671.6 \relax& \protect[S~\textsc{ii}] \relax& $39 \pm 6$ \relax& \relax& $61 \pm 7$ \relax& $41 \pm 9$ \relax& $60 \pm 10$ \relax& $89 \pm 12$  \\
673.1 \relax& \protect[S~\textsc{ii}] \relax& $116 \pm 7$ \relax& \relax& $133 \pm 6$ \relax& $157 \pm 9$ \relax& $143 \pm 13$ \relax& $121 \pm 12$  \\
732.0 \relax& \protect[O~\textsc{ii}] \relax& $58 \pm 6$ \relax& \relax& $48 \pm 6$ \relax& $59 \pm 8$ \relax& $53 \pm 10$ \relax& $22 \pm 16$  \\
733.0 \relax& \protect[O~\textsc{ii}] \relax& $61 \pm 6$ \relax& \relax& $47 \pm 6$ \relax& $60 \pm 10$ \relax& $76 \pm 15$ \relax& $53 \pm 15$  \\
751.5 \relax& \protect[Na~\textsc{i}] \relax& \relax& \relax& $< 30$ \relax& $< 40$ \relax& $< 50$ \relax& $< 70$  \\
766.4\textsuperscript{\scriptsize{b}} \relax& K~\textsc{i} \relax& $680 \pm 30$ \relax& \relax& $982 \pm 19$ \relax& $1240 \pm 30$ \relax& $920 \pm 30$ \relax& $1060 \pm 40$  \\
772.5 \relax& \protect[S~\textsc{i}] \relax& $333 \pm 7$ \relax& \relax& $361 \pm 7$ \relax& $411 \pm 12$ \relax& $351 \pm 15$ \relax& $371 \pm 16$  \\
777.4 \relax& O~\textsc{i} \relax& $158 \pm 7$ \relax& \relax& $166 \pm 7$ \relax& $317 \pm 11$ \relax& $190 \pm 13$ \relax& $269 \pm 14$  \\
818.3 \relax& Na~\textsc{i} \relax& {---}\relax& \relax& $117 \pm 19$ \relax& $410 \pm 30$ \relax& $210 \pm 30$ \relax& $90 \pm 20$  \\
819.5 \relax& Na~\textsc{i} \relax& {---}\relax& \relax& $360 \pm 20$ \relax& $840 \pm 30$ \relax& $350 \pm 30$ \relax& $240 \pm 40$  \\
844.6 \relax& O~\textsc{i} \relax& $196 \pm 7$ \relax& \relax& $266 \pm 9$ \relax& $400 \pm 15$ \relax& $281 \pm 17$ \relax& $250 \pm 20$  \\
922.3 \relax& S~\textsc{i} \relax& $392 \pm 10$ \relax& \relax& {---}\relax& {---}\relax& {---}\relax& {---} \\
\bottomrule
\multicolumn{8}{p{0.85\textwidth}}{\textbf{Notes.} Numbers preceded by a less-than symbol ($<$) are 2$\sigma$ upper limits for non-detections. Uncertainties are representative of photon counting (Poisson) noise and do not capture additional systematic uncertainties (which are especially present for the 589.0 and 589.6~nm \NaI{} doublet). Blank cells indicate wavelengths not captured by choice of cross-disperser and combination of cross-disperser and echelle angles. Order overlap prevented characterization and subtraction of backgrounds for wavelengths below approximately 380~nm. Background subtraction failed for Na\,\textsc{i} 818.3 and 819.5~nm on 2022 November 24. With the exception of 2022 November 24, fringing prevented proper background subtraction from all S\,\textsc{i} 922.3~nm data, though the emission was still visibly present in all spectra.}\\[-5pt]
\multicolumn{8}{p{0.85\textwidth}}{
\begin{itemize}[nosep, labelsep=-1.5pt, align=left, leftmargin=*]
\item[\textsuperscript{a}]First detected in eclipse by \citet{Bouchez2000}.
\item[\textsuperscript{b}]First detected in eclipse by \citet{Schmidt2023}.
\end{itemize}
}
\end{tabular}
\end{table*}

\begin{figure*}
    \centering
    \includegraphics[width=\textwidth]{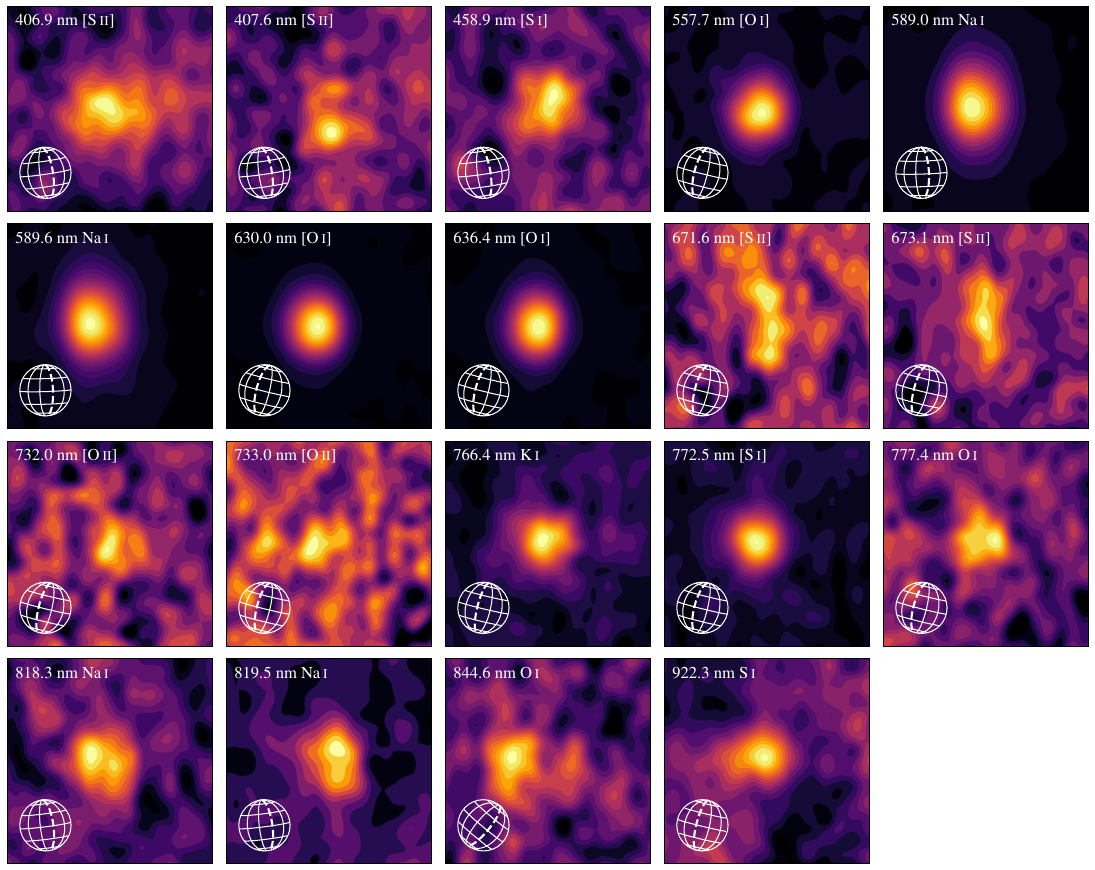}
    \caption{Calibrated example images for all 19 detected auroral emissions listed in Table \ref{tab:all-lines} displayed using 20 discrete contours. Each image is a single, five-minute integration with the background subtracted. To better reveal the emissions, we smoothed the data using a two-dimensional Gaussian kernel with an FWHM of \ang{;;0.5} approximating typical seeing conditions for Maunakea. Due to the large range of brightnesses, we scaled each image using its individual dynamic range, so comparing apparent brightnesses between images is not meaningful. Graticules in the lower right of each panel show the size and orientation of Io during the observation; the thicker dashed line shows the location of the prime meridian. (We chose not to display these directly over the emission in order to not interfere with interpretation of the dimmer emissions and to prevent the need to display three overlapping globes for the 844.6~nm \OI{} triplet.) For the 777.4~nm \OI{} and 922.3~nm \SI{} triplets we have displayed only the brightest, shortest wavelength component due to the large separation of the three emissions on the HIRES detector. However, we used all three components simultaneously when retrieving the total brightnesses. The three components of the 844.6~nm \OI{} triplet are somewhat blended at the HIRES detector resolution, so we centered that images on its average wavelength. The 732.0 and 733.0 [\OII{}] emissions are both doublets and we chose to the brighter components at 731.999 and 732.967~nm, respectively. \\[4pt] (The data used to create this figure are available.)}
    \label{fig:data}
\end{figure*}

Our analysis of the HIRES spectra revealed emission lines from a variety of neutral and singly-ionized atoms including O, \ce{O+}, Na, S, \ce{S+} and K. This complete set of emissions includes the first Io eclipse detections of the \OI{} lines at 777.4 and 844.6~nm, the \NaI{} lines at 818.3 and 819.5~nm, the [\SI{}] lines at 458.9 and 772.5~nm, the \SI{} triplet at 922.3~nm, the [\OII{}] lines at 732.0 and 733.0~nm and the [\SII{}] lines at 406.9, 407.6, 671.6 and 673.1~nm. Figure \ref{fig:data} shows examples of what these emissions (along with the previously identified emissions) look like in the reduced HIRES detector images. Io's angular width was nearly that of the slit (see Figure \ref{fig:viewing-geometry}), so any emission from an extended corona would appear only in the vertical direction in the detector images.

Table \ref{tab:all-lines} lists each of the emission lines for which we attempted to detect auroral emission along with the average brightness for those found at a signal-to-noise of 2 or greater. For non-detections (negative brightness or signal-to-noise below 2) we have instead reported the 2$\sigma$ upper limit for the brightness. Brightnesses are in units of rayleighs (R), defined as 
\begin{equation}\label{eq:rayleigh}
    1~\mathrm{R} \equiv \frac{10^{10}}{4\mathrm{\pi}}~\mathrm{ph~s^{-1}~m^{-2}~sr^{-1}}.
\end{equation}
When comparing to brightnesses published by others, we occasionally used kilorayleighs (kR), where $1~\mathrm{kR}=1000~\mathrm{R}$. Significant order overlap prevented us from calculating any emission brightness for wavelengths shorter than about 380~nm. Additionally, for our choice of echelle and cross disperser angles, the \KI{} emission at 769.9~nm did not fall onto the detector, though it should have been at a detectable brightness given the previously observed brightness ratio between the potassium D lines \citep{Schmidt2023}.

Because HIRES cannot operate with both HIRESr and HIRESb on the same night, we were unable to simultaneously observe the shorter and longer wavelength emissions. Additionally, the inherent temporal variability in the total brightnesses precludes meaningful comparisons. However, when we compared the relative brightness of the [\OI{}] 557.7~nm emission line and the [\SI{}] electronic equivalent at 772.5~nm, we found they had a Pearson correlation coefficient of 0.917 with a \textit{p}-value of 0.0282, which qualifies as statistically significant beyond a standard 95\% confidence threshold. We also compared the relative brightness of the [\OI{}] 557.7~nm emission line to the [\SII{}] 673.1~nm emission line and found they had a Pearson correlation coefficient of 0.813 with a \textit{p}-value of 0.0942, statistically significant for a 90\% confidence threshold.

This indicates that the absolute brightnesses of the sulfur and oxygen emissions mostly co-vary and therefore the brightnesses and uncertainties derived from HIRESr observations can be scaled to approximate their expected values on 2023 August 09. Since we were able to retrieve the 557.7, 630.0 and 636.4~nm [\OI{}] brightnesses with both cross dispersers, we used the relative weighted average brightness of these three oxygen emissions to calculate scaling factors and uncertainties of $1.1 \pm 0.2$ for 2022 November 24, $0.92 \pm 0.17$ for 2023 August 25, $0.83 \pm 0.15$ for 2024 September 12, $0.73 \pm 0.12$ for 2024 October 05 and $0.88 \pm 0.16$ for 2024 October 21.

The brightnesses we retrieved from our HIRES observations vary substantially from those reported by \citet{Bouchez2000}, potentially indicating large secular variability in aurora brightness. They found a 557.7~nm [\OI{}] brightness of $1.3 \pm 0.2$~kR, a factor of $3.0 \pm 0.5$ larger than the weighted average brightness of $0.429 \pm 0.004$~kR averaged over the six nights in our data. We found a similar factor of $3.0 \pm 0.4$ for the comparison of the 630.0~nm [\OI{}] brightness, and a slightly larger factor of $3.2 \pm 0.5$ for the 636.4~nm [\OI{}] brightness. However, for the sodium doublet, their results were dimmer by a factor of $0.28 \pm 0.04$ for the 589.0~nm \NaI{} D$_1$ line and $0.29 \pm 0.04$ for the 589.6~nm \NaI{} D$_2$ line. These differences point to stochastic variability in atmospheric column density (discussed further in Section \ref{sec:electron-scale-height}).

The weighted average brightness we found for the \ce{O(^1D)} doublet was $7.75 \pm 0.02$~kR, which exceeds the 4.8~kR average reported by \citet{Schmidt2023} despite cross-calibration of our two analysis routines (our pipeline and theirs produced comparable brightnesses when reducing the same HIRES data). This suggests the difference is a real change in brightness and not a systematic difference in our respective flux calibrations. The eclipsed \ce{O(^1D)} doublet brightness is less than the 10.5~kR average reported in the sunlit measurements by \citet{Oliversen2001}. We found no evidence for time-dependence of the O emissions during the eclipse phase, while the Na emissions do show a strong time dependence due to the sudden loss of solar photochemical production pathways \citep{Schmidt2023}. Both \citet{Oliversen2001} and \citet{Schmidt2023} found weak correlations between \ce{O(^1D)} emission and Io's location relative to the plasma torus, but they observed considerable variance by up to a factor of 3 in the brightness at a given magnetic longitude.

\subsection{Comparison with Cassini/ISS Images}
\label{sec:cassini-comparison}

The identification of so many discrete emission features provided the opportunity to better understand Io's species-dependent auroral morphology by comparing with images of Io taken in eclipse, first done by \citet{Bouchez2000} when they compared their five detected emission lines to images of Io in eclipse taken by the Solid State Imaging experiment (SSI) on the Galileo Orbiter spacecraft \citep{Geissler1999}.

\begin{figure}
    \centering
    \includegraphics[width=\columnwidth]{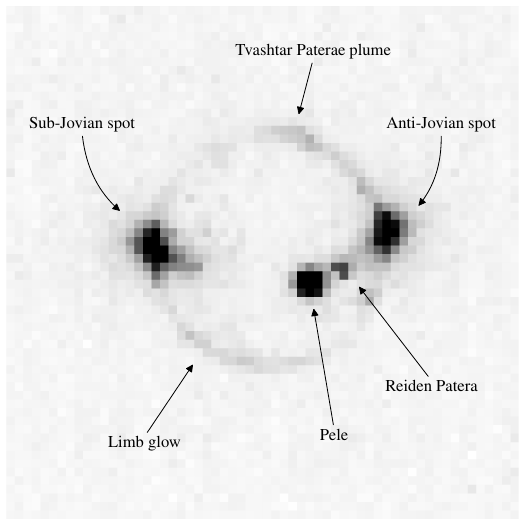}
    \caption{Morphological features of Io's optical emission in eclipse. This image was taken by the Narrow Angle Camera (NAC) on Cassini's Imaging Science Subsystem (ISS) with the CL1+CL2 (clear) filter combination, so it includes emission across the full detector sensitivity range from approximately 235 to 1100~nm. The volcanic features Pele and Reiden Patera also include thermal emission, especially at longer wavelengths.} Table \ref{tab:app-cassini-image-filenames} lists the file name and observation time for this image.\\[4pt] (The data used to create this figure are available.)
    \label{fig:cassini-morphology}
\end{figure}

\begin{figure}
    \centering
    \includegraphics[width=\columnwidth]{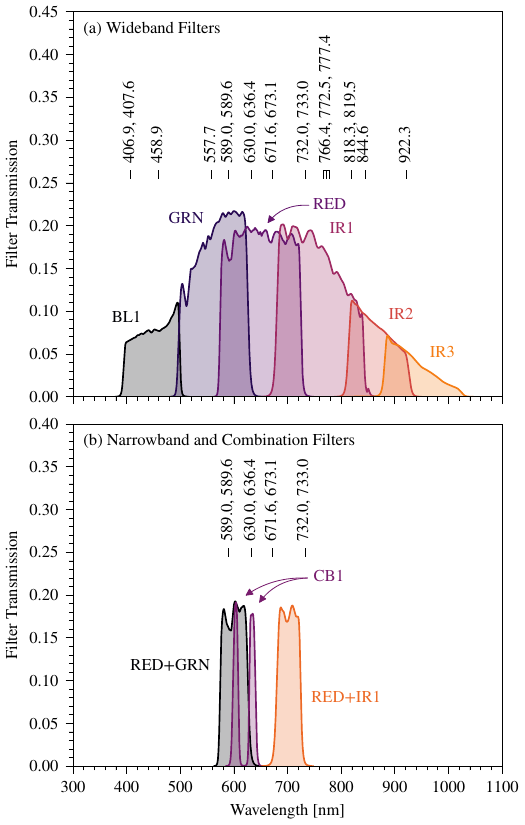}
    \caption{Cassini/ISS NAC filter transmission \citep{Porco2004} and locations of identified auroral emission features. (a) shows all the wideband filters and the locations of all auroral lines, grouped as necessary to avoid overlap of the labels. (b) shows the narrowband filter CB1 and the filters made by the combination of two broadband filters, along with the emission features which correspond to those filter bandpasses.\\[4pt] (The data used to create this figure are available.)}
    \label{fig:cassini-filters}
\end{figure}

\begin{figure}
    \centering
    \includegraphics[width=\columnwidth]{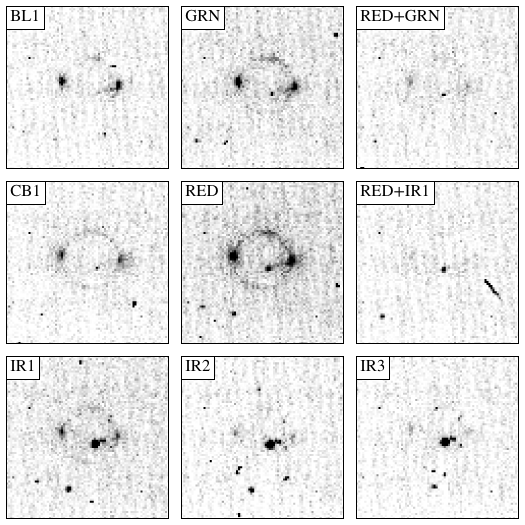}
    \caption{Images of each Cassini/ISS filter and filter combination containing discrete auroral emissions identified in the HIRES spectra. Labels in the upper right indicate the filter or filter combination. Table \ref{tab:app-cassini-image-filenames} lists the individual file names and observation time for the images displayed for each filter.\\[4pt] (The data used to create this figure are available.)}
    \label{fig:cassini-filter-images}
\end{figure}

\begin{figure}
    \centering
    \includegraphics[width=\columnwidth]{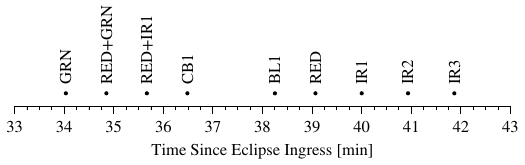}
    \caption{Cassini/ISS imaging filter sequence displayed in Figure \ref{fig:cassini-filter-images} sorted by the exposure midpoint time (see Table \ref{tab:app-cassini-image-filenames}) of each image relative to eclipse ingress, which occurred at 2001 January 05 11:03 UTC.}
    \label{fig:cassini-images-times}
\end{figure}

On its way to Saturn, the Narrow Angle Camera (NAC) of Cassini’s Imaging Science Subsystem \citep[ISS,][]{Porco2004} took both broadband and narrowband images of Io in eclipse during its flyby of the Jovian system (see Figure \ref{fig:cassini-morphology} for a broadband clear filter example image and Figure \ref{fig:cassini-filters} for a description of the filter bandpasses). ISS imaged Io's trailing hemisphere, in contrast to the HIRES eclipse observing geometry which is restricted to the sub-jovian hemisphere. Figure 5 of \citet{Geissler2004} shows images taken under 15 different filters and filter combinations on 2001 January 05, nine of which (shown in our Figure \ref{fig:cassini-filter-images}) had bandpasses which contained wavelengths of the auroral emissions we detected in the HIRES spectra. ISS took four sequences of images with exposure times of 12 seconds each. The sequences took about 14 minutes and 22 seconds to complete and they began and ended with clear filter images. We chose to display images taken during the second sequence in Figure \ref{fig:cassini-filter-images} due to their high signal-to-noise. Figure \ref{fig:cassini-images-times} shows the filter sequence sorted by the exposure midpoint time of each image relative to eclipse ingress, which occurred at 2001 January 05 11:03 UTC.

These images demonstrate that Io's optical aurora exhibit some combination of two primary morphological features: the sub/anti-Jovian spots (also called the ``equatorial glow'') which were first identified in UV imaging \citep[e.g.,][]{Retherford2000,Retherford2003,Retherford2007,Roesler1999,Roth2011,Roth2014,Roth2017,Saur2000} and a diffuse limb glow \citep[e.g.,][]{Retherford2003,Moore2010}. Because the HIRES data do not have the spatial resolution to resolve spots versus a diffuse equatorial enhancement distributed across all visible longitudes, we used the term ``equatorial glow'' throughout our analysis instead of sub-Jovian spot when referring to emission features in the HIRES data (we still used the term for the ISS images). Further, despite its name, the sub-Jovian spot actually appears up to \ang{20} west (wakeward) of the sub-Jovian longitude \citep{Retherford2007,Moore2010}. Regardless, the FWHM of the bright emissions in Figure \ref{fig:data} are smaller than the angular size of Io's disk (indicated by the graticules in the lower left corners of each image), so the observed emissions are spatially isolated and likely to be primarily from the sub-Jovian spot.

The aurora also show transient thermal and/or thermally-excited emissions from discrete volcanic features including Reiden Patera, the volcano Pele and a spot near the north pole which they identify as being associated with the Tvashtar Paterae volcanic plume. Reiden Patera and Pele appear in the CLR (clear) filter image (a combination of filters CL1 and CL2) and are increasingly more prominent at longer wavelengths due to their thermal emission. The Tvashtar Paterae plume appears under most filters. Figure \ref{fig:cassini-morphology} shows an example ISS CL1+CL2 (clear) filter combination image containing each of these emission features, while Figure \ref{fig:cassini-filters} shows the transmission bandpasses of the nine filters and the locations of the auroral emissions identified in the HIRES spectra.

\begin{table*}
    \centering
    \caption{Emission Morphologies in Cassini/ISS Images and Contributing Auroral Emissions Identified in HIRES Spectra}
    \label{tab:cassini-filters}
    \begin{tabular}{lclllccc}
    \toprule
     & & & & & \multicolumn{3}{c}{Disk-Integrated Brightness} \\
     \cmidrule(lr){6-8}
    Filter & Bandpass\textsuperscript{\scriptsize{a}} [nm] & Equatorial Glow & Limb Glow & Plume & HIRES\textsuperscript{\scriptsize{b}} [kR] & ISS\textsuperscript{\scriptsize{c}} [kR] & $\mathrm{HIRES/ISS}$ Ratio\\
    \midrule
    BL1 & 390 to 500 & [\SI{}], [\SII{}] & [\SI{}]? & [\SII{}]? & $0.50 \pm 0.02$ & $25 \pm 10$ & $0.020 \pm 0.008$ \\
    GRN & 495 to 635 & [\OI{}], \NaI{} & [\OI{}] & \NaI{} & $24.71 \pm 0.09$ & $15 \pm 7$ & $1.6 \pm 0.8$ \\
    RED+GRN & 570 to 635 & \NaI{} & --- & \NaI{} & $24.28 \pm 0.09$ & $4 \pm 5$ & $6 \pm 8$ \\
    CB1 & 595 to 615, 625 to 645 & [\OI{}] & [\OI{}] & --- & $7.845 \pm 0.019$ & $11 \pm 4$ & $0.7 \pm 0.3$ \\
    RED & 570 to 730 & [\OI{}], \NaI{}, [\SII{}] & [\OI{}] & \NaI{}, [\SII{}] & $32.31 \pm 0.09$ & $19.1 \pm 0.7$ & $1.69 \pm 0.06$\\
    RED+IR1 & 670 to 730 & [\OII{}], [\SII{}] & --- & [\OII{}], [\SII{}] & $0.293 \pm 0.007$ & --- & --- \\
    IR1 & 670 to 850 & \OI{}, [\OII{}], \NaI{}, [\SI{}], \KI{} & \OI{}? [\SI{}]? & [\OII{}]?, \NaI{}? & $2.49 \pm 0.02$ & $11 \pm 3$ & $0.23 \pm 0.06$\\
    IR2 & 800 to 940 & \OI{}, \NaI{}, \SI{} & \OI{}, \SI{}? & \NaI{}? & $0.856 \pm 0.019$ & --- & --- \\
    IR3 & 880 to 1025 & \SI{} & \SI{}? & --- & $0.392 \pm 0.010$ & --- & --- \\
    \bottomrule
    \multicolumn{8}{p{\textwidth}}{
    \textbf{Note.} Question marks indicate assumed contributions inferred but not unambiguously identified with the combination of ISS images and HIRES spectra.
    \begin{itemize}[nosep, labelsep=-1.5pt, align=left, leftmargin=*]
    \item[\textsuperscript{a}]From Table 1 of \citet{Geissler2004}.
    \item[\textsuperscript{b}]Weighted average of brightnesses from Table \ref{tab:all-lines}.
    \item[\textsuperscript{c}]Converted from Table 4 of \citet{Geissler2004} as described in Section \ref{sec:cassini-comparison} of this paper.
    \end{itemize}}
    \end{tabular}
\end{table*}

\citet{Geissler2004} reported the brightnesses of specific auroral morphologies like the anti-Jovian equatorial glow (their Tables 1 and 5) and the limb glow (their Table 3). They did not report disk-integrated brightnesses, which make comparisons with the disk-integrated HIRES brightnesses difficult. However, they did report disk-integrated total photon fluxes and uncertainties in their Table 4. They calculated these fluxes over apertures which included all apparent emission: the equatorial spots, the limb glow and any potential extended corona, and converted them to total photon flux emitted into $4\mathrm{\pi}~\mathrm{sr}$ around the entire satellite. We took these flux values (they did not specify units, so we assumed $\mathrm{ph~s^{-1}}$), divided by the cross-sectional area of Io ($1.0514\times 10^{12}~\mathrm{m}^2$) and $4\mathrm{\pi}~\mathrm{sr}$, then converted to kR using Equation \eqref{eq:rayleigh}.
We have included these brightnesses in Table \ref{tab:cassini-filters}.

The BL1 filter image brightness is significantly larger than that derived from the HIRES data. \citet{Geissler2004} attributed the brightness in the BL1 filter image to electron-impact quasi-continuum fluorescence of \ce{SO2}, though they also proposed some contribution from the \ce{S2(B $\mathrm{^3\Sigma^-_g}$ -> X $\mathrm{^3\Sigma^-_u}$)} band system to account for the larger-than-expected brightness in this filter relative to the UV1 filter. \citet{Trafton2012} also observed molecular emission bands in this wavelength region from Io in eclipse.

The GRN filter image contains both the 557.7~nm [\OI{}] and 589.0 and 589.6~nm \NaI{} doublet emissions, while the RED+GRN combination filter contains just the \NaI{} doublet emissions. These images were taken only about 1 minute apart (see Figure \ref{fig:cassini-images-times}), so the fact that they recorded nearly identical brightnesses suggests that the [\OI{}] emission dominates in the GRN filter.

The CB1 filter image contains just the 630.0 and 636.4~nm [\OI{}] emissions, which shows that the oxygen emission primarily comes from the equatorial and limb glows and does not seem to be correlated with the volcanic plume. This conclusion is further supported by the RED filter image, which contains the 589.0 and 589.6~nm \NaI{} doublet, the 630.0 and 636.4~nm [\OI{}] lines and the 671.6 and 673.1~nm [\SII{}] lines. The strong plume emission must come from the \NaI{} doublet and [\SII{}] lines, suggesting that the plume emission in the GRN filter is probably from the \NaI{} doublet and not from the 557.7~nm [\OI{}] line, though the different lifetimes between the \ce{O(^1S)} and \ce{O(^1D)} states could lead to preferential quenching of the 630.0 and 636.4~nm [\OI{}] lines in the high density plume. The RED+IR1 filter combination image contains the 671.6 and 673.1~nm [\SII{}]  and 732.0 and 733.0~nm [\OII{}] emissions, but they fall near the edges of the bandpass, accounting for the nearly featureless image.

The limb glow in the IR1 filter is likely from the \OI{} 777.4 and 844.6~nm emissions, and perhaps also from the 772.5~nm [\SI{}] and line (the sulfur equivalent to the 557.7~nm [\OI{}] line). \citet{Geissler2004} proposed the largest contribution to the total brightness in the IR1 filter came from the 766.4~nm \KI{} line, which agrees with the HIRES data, where it accounted for about half of the total disk-integrated brightness across the filter's bandpass. The IR2 and IR3 filters contain the emissions identified in the HIRES spectra at wavelengths greater than 800~nm, but the signal-to-noise is quite low, and only the equatorial glows are significant.

Table \ref{tab:cassini-filters} summarizes the species we attribute to the observed morphologies in each filter. Overall, the equatorial glow seems to include emission from all detected species, both neutrals and ions. The limb glow seems to be associated with the bright neutral oxygen emissions, though we cannot rule out contribution from neutral sulfur as well, which may be the species contributing to the extremely faint limb glows in the BL1, IR2 and IR3 filters. The concentration of molecular species at low-to-mid latitudes \citep{Strobel2001} and global atomic coronae \citep{Wolven2001} suggest dissociative electron impact on molecules like \ce{SO2} and \ce{SO} produce the bright auroral spots while excitation of the atomic S and O coronae produce the limb glow. Emission from the Tvashtar plume is probably primarily from sodium but may also include emission from ionized atomic oxygen and sulfur.

\section{Results and Discussion}

\subsection{Electron Energy}
\label{sec:electron-energies}

\begin{figure}
    \centering
    \includegraphics[width=\linewidth]{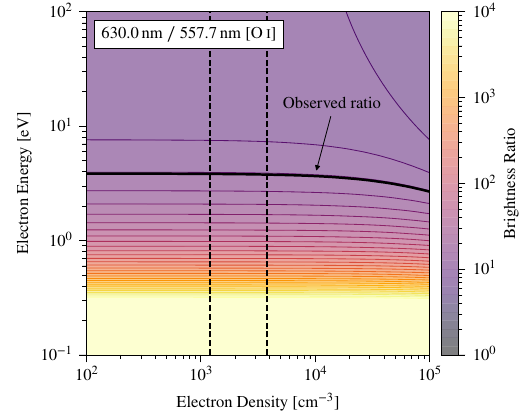}
    \caption{Dependence of the $630.0~\mathrm{nm}/557.7~\mathrm{nm}$ [\OI{}] brightness ratio on electron energy and density assuming electron impact on an atmospheric column of atomic oxygen. The dashed vertical black lines show the typical electron density range of 1200 to 3800~$\mathrm{cm}^{-3}$ experienced by Io through one rotation of Jupiter's magnetosphere \citep{Bagenal2020}. The thick black line shows our average observed ratio of $13.64 \pm 0.13$. This brightness ratio for electron impact on atomic oxygen is effectively insensitive to the range of typical electron densities experienced by Io during one rotation of Jupiter's magnetic field, making it diagnostic of the energy of the exciting electrons. However, electron impact on atomic oxygen cannot account for the observed \OI{} 777.4 and 844.6~nm brightnesses. Colored lines highlight contours at a spacing of 0.1~dex. Calculated using CHIANTI v10.1 \citep{Dere1997,Dere2023}.\\[4pt] (The data used to create this figure are available.)}
    \label{fig:6300-5577-electron-energy}
\end{figure}

Emission line ratios are sensitive to different electron energy distributions because of their individual energy-dependent excitation cross sections. In contrast, absolute auroral brightnesses are sensitive to both electron energy distributions and the product of electron number densities and atmospheric column densities. \citet{Schmidt2023} evaluated the $630.0~\mathrm{nm}/557.7~\mathrm{nm}$ [\OI{}] brightness ratio in observations taken primarily with the ARC Echelle Spectrograph (ARCES) at the Apache Point Observatory and found it was higher than expected from electron impact on a column of just atomic oxygen. They attributed the larger ratio to electrons losing energy as they precipitated through Io's \ce{SO2} atmosphere, allowing them to preferentially excite the 630.0~nm red line emission (which has a lower excitation energy than the 557.7~nm emission). Models like those of \citet{Saur1999} and \citet{Dols2012} show electron energies decreasing from a typical upstream torus energy of 5~eV down to 0.2~eV near Io's surface. Figure \ref{fig:6300-5577-electron-energy} shows this brightness ratio for electron impact on atomic oxygen as a function of electron energy and density. For electron impact on O the emission line ratio is sensitive to energy but independent of plasma density for the range of values plausible for Io. Though the impact of 3 to 4~eV electrons on a column of atomic oxygen can explain the observed $630.0~\mathrm{nm}/557.7~\mathrm{nm}$ [\OI{}] brightness ratios (Figure \ref{fig:6300-5577-electron-energy} shows the average ratio of $13.64 \pm 0.13$), it cannot account for the ratios with the other oxygen emissions (777.4 and 844.6~nm) detected in the HIRES spectra. Quenching and electron energy variation with latitude, altitude and within plumes is a plausible alternative explanation to the one we will present in Section \ref{sec:parent-species}.

\subsection{Auroral Parent Species}
\label{sec:parent-species}

In order to explore how the emissions change with other possible atmospheric compositions, we compared brightness calculations from our auroral emission model for three combinations: \ce{O} alone (simulating emission from just the atomic oxygen column), an atmosphere of \ce{O} and \ce{SO2}, and an atmosphere of \ce{O}, \ce{SO2} and a third molecular component. Because of the limited cross sections available for \ce{SO2}, we were only able to evaluate three oxygen emissions: [\OI{}] at 557.7 and \OI{} at 777.4 and 844.6~nm. We used a Maxwellian electron energy distribution centered at 5~eV to determine if the observed brightness ratios could be explained through impact by the upstream electron population. Table \ref{tab:modeled-emissions} lists the results produced by our aurora model for the five different atmospheric compositions discussed below. 

\begin{figure}
    \centering
    \includegraphics[width=\columnwidth]{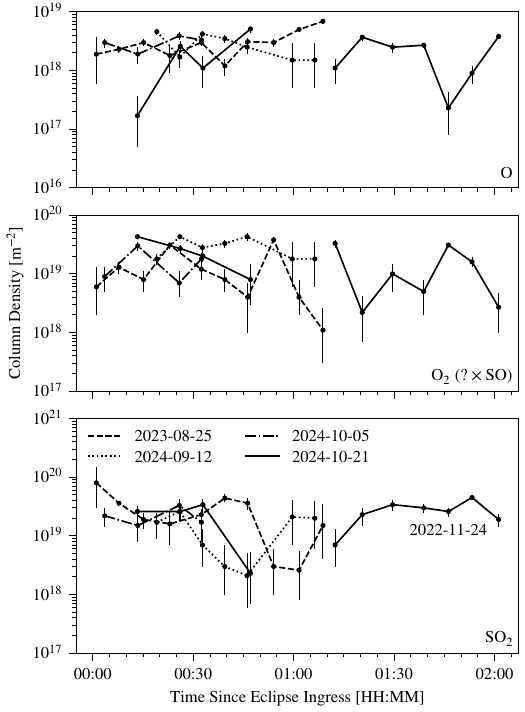}
    \caption{Best-fit emission model column densities for O (top), \ce{O2} as a proxy for \ce{SO} (middle) and \ce{SO2} (bottom). These results assume only direct excitation of O and dissociative excitation of \ce{SO} and \ce{SO2} and do not include any recombination processes. We used a Maxwellian electron distribution centered at 5~eV with a density of $\mathrm{2500~cm^{-3}}$. Line styles differentiate between the five observation nights as notated in the bottom plot. The variance in the \ce{O2} column densities are physically meaningful, but the absolute magnitude is not necessarily representative of the \ce{SO} column density (see the discussion in Section \ref{sec:electron-energies}).\\[4pt] (The data used to create this figure are available.)}
    \label{fig:column-density-timeseries}
\end{figure}

The single-component atmosphere comprised of just \ce{O} alone cannot replicate the observed brightness ratios of the three emission lines: it under-predicts the 777.4~nm brightnesses and over-predicts the 557.7 and 844.6~nm brightnesses. Including \ce{SO2} improves the fit overall, but still fails to reproduce the observed brightnesses within their respective uncertainties, resulting in an over-prediction of the 844.6~nm emission and an under-prediction of the 777.4~nm emission. However, including \ce{O2} results in an excellent fit for all three emission features well within their observed uncertainties. However, this result yields \ce{O2} column densities of approximately $10^{19}$~cm$^{-2}$ (see Figure \ref{fig:column-density-timeseries}), which is about 2 orders of magnitude larger than the estimated 1\% mixing ratio between \ce{O2} and \ce{SO2} \citep{Moses2002}. This probably means the observed auroral emissions are from a combination of electron impact on exospheric columns of \ce{O}, \ce{SO2} and a molecule that behaves like \ce{O2} when dissociated into excited fragments by electron impact. The most likely candidate molecule is SO, which has a mixing ratio with \ce{SO2} of around 10\% \citep{McGrath2000} and may be higher during eclipse when the \ce{SO2} atmosphere condenses onto the surface \citep{Tsang2016}. Unfortunately, emission and/or excitation cross sections have not yet been published for \ce{SO} \citep{McConkey2008}, so it cannot be explicitly included in our auroral emission model. 

S and O occupy the same group in the periodic table, the valence electrons in the p orbitals are the same for both molecules, and therefore their overall state distribution is similar. They also have comparable dissociation energies: approximately 5.1~eV for \ce{O2} and 5.4~eV for \ce{SO} \citep{Darwent1970}. Theoretical calculations show similar excitation cross sections between \ce{O2} and \ce{SO}, though the \ce{SO} cross sections tend to be larger due to the physical size difference between the two molecules \citep{Rajvanshi2010}. We therefore chose to use the \ce{O2} cross sections as a proxy for the \ce{SO} cross sections. However, while the relative cross sections (and therefore the brightness ratios) between oxygen emission lines will be similar, the absolute cross sections may not be of the same magnitude. Consequently, while the modeled component from \ce{O2} probably includes emission primarily from \ce{SO} (along with any trace \ce{O2} present), we cannot determine their relative contributions due to the inherent degeneracy. As a result, while the O and \ce{SO2} column outputs are real, only the relative variability within the \ce{O2} column density is real; the absolute column density magnitude is not physically meaningful, and should not be interpreted as representative of the either the modeled \ce{O2} or \ce{SO} column density. The true \ce{SO} column density is likely smaller than the modeled result if the \ce{SO} cross sections are systematically larger than the \ce{O2} cross sections \citep{Rajvanshi2010}.

\begin{table*}
    \centering
    \caption{Best-Fit Auroral Model Oxygen Emissions for Different Atmospheric Compositions}
    \label{tab:modeled-emissions}
    \begin{tabular}{lcS[table-format=3.0,detect-all]S[table-format=3.0,detect-all]S[table-format=3.0,detect-all]S[table-format=3.0,detect-all]S[table-format=3.0,detect-all]}
\toprule
 \relax& \relax& \multicolumn{5}{c}{Model Atmosphere Composition}\\
\cmidrule{3-7}
Date \relax& Observed \relax& {\ce{O}} \relax& {$\ce{O} + \ce{SO2}$} \relax& {$\ce{O} + \ce{SO2} + \ce{O2}$} \relax& {$\ce{O} + \ce{SO2} + \ce{CO2}$} \relax& {$\ce{O} + \ce{SO2} + \ce{H2O}$}\\
 \hspace*{{1em}} (Emission) \relax& {[R]} \relax& {[R]} \relax& {[R]} \relax& {[R]} \relax& {[R]} \relax& {[R]}\\
\midrule
\textbf{2022 November 24}\\
\hspace*{1em}557.7~nm \relax& $395 \pm 7$ \relax& 327 \relax& \bfseries 399 \relax& \bfseries 395 \relax& \bfseries 401 \relax& \bfseries 400\\
\hspace*{1em}777.4~nm \relax& $158 \pm 7$ \relax& 102 \relax& 88 \relax& \bfseries 157 \relax& 88 \relax& 87\\
\hspace*{1em}844.6~nm \relax& $196 \pm 7$ \relax& 300 \relax& 214 \relax& \bfseries 196 \relax& 215 \relax& 214\\
\textbf{2023 August 25}\\
\hspace*{1em}557.7~nm \relax& $417 \pm 7$ \relax& 385 \relax& \bfseries 423 \relax& \bfseries 415 \relax& \bfseries 420 \relax& \bfseries 419\\
\hspace*{1em}777.4~nm \relax& $166 \pm 7$ \relax& 120 \relax& 109 \relax& \bfseries 166 \relax& 108 \relax& 108\\
\hspace*{1em}844.6~nm \relax& $266 \pm 9$ \relax& 353 \relax& 292 \relax& \bfseries 266 \relax& 293 \relax& 293\\
\textbf{2024 September 12}\\
\hspace*{1em}557.7~nm \relax& $481 \pm 10$ \relax& \bfseries 498 \relax& \bfseries 500 \relax& \bfseries 483 \relax& \bfseries 498 \relax& \bfseries 499\\
\hspace*{1em}777.4~nm \relax& $317 \pm 11$ \relax& 155 \relax& 154 \relax& \bfseries 312 \relax& 153 \relax& 153\\
\hspace*{1em}844.6~nm \relax& $400 \pm 15$ \relax& 457 \relax& 451 \relax& \bfseries 381 \relax& 448 \relax& 449\\
\textbf{2024 October 05}\\
\hspace*{1em}557.7~nm \relax& $440 \pm 12$ \relax& 412 \relax& \bfseries 448 \relax& \bfseries 437 \relax& \bfseries 438 \relax& \bfseries 441\\
\hspace*{1em}777.4~nm \relax& $190 \pm 13$ \relax& 128 \relax& 116 \relax& \bfseries 190 \relax& 113 \relax& 114\\
\hspace*{1em}844.6~nm \relax& $281 \pm 17$ \relax& 378 \relax& \bfseries 315 \relax& \bfseries 280 \relax& \bfseries 312 \relax& \bfseries 314\\
\textbf{2024 October 21}\\
\hspace*{1em}557.7~nm \relax& $444 \pm 12$ \relax& \bfseries 429 \relax& \bfseries 457 \relax& \bfseries 443 \relax& \bfseries 451 \relax& \bfseries 452\\
\hspace*{1em}777.4~nm \relax& $269 \pm 14$ \relax& 134 \relax& 121 \relax& \bfseries 271 \relax& 119 \relax& 119\\
\hspace*{1em}844.6~nm \relax& $250 \pm 20$ \relax& 394 \relax& 328 \relax& \bfseries 251 \relax& 327 \relax& 328\\
\bottomrule
\multicolumn{7}{p{0.7\textwidth}}{\textbf{Note.} Bold numbers indicate model brightness values within $2\sigma$ of the observed brightness.}
\end{tabular}
\end{table*}

To ensure that this exceptionally good agreement from the inclusion of the \ce{O2} component wasn't an artifact from fitting three emission lines with a three species atmosphere, we fit two additional three-species atmospheres, substituting \ce{H2O} and \ce{CO2} (the only other species for which we had 557.7, 777.4 and 844.6~nm emission cross sections) for the \ce{O2} component. The other three-component atmospheres always resulted in a poor fit for at least one emission line. This could mean that the auroral emission is confined to high altitudes where electrons still maintain their 5~eV plasma torus energies. However, the Cassini/ISS images (specifically the CB1 image in Figure \ref{fig:cassini-filter-images}) seem to show the longest lifetime 630.0 and 636.4~nm emissions all the way down to Io's surface. So, while the observed oxygen aurora brightness ratios can be explained by electron impact on Io's \ce{O}, \ce{SO2} and \ce{SO} exospheres by 5~eV torus electrons without invoking the collisional cooling necessary for emission from a pure \ce{O} column proposed by \citet{Schmidt2023}, electron energies are much lower in the denser parts of the ionosphere near to Io's surface \citep{Saur1999,Dols2008,Dols2012,Dols2024} and therefore there exists a degeneracy between electron energy and composition which cannot be resolved using forbidden oxygen emissions from Io.

Regardless, the disk-integrated 557.7, 777.4 and 844.6~nm brightnesses are consistent with electron impact on an atmosphere composed of O, \ce{SO} and \ce{SO2}. For the 557.7~nm [\OI{}] line, the fractional contribution to the total brightness from the \ce{SO} component varies between 6 and 9\%.

\subsection{Atmosphere Collapse in Eclipse}

It remains unclear what fraction of Io's \ce{SO2} atmosphere comes from volcanic outgassing and what fraction comes from sublimation of surface ice. Some observations show a large decrease in column density during eclipse associated with the rapid drop in surface temperature, suggesting a predominantly sublimation-supported atmosphere where solar insolation maintains the vapor-pressure equilibrium between Io's \ce{SO2} atmosphere and surface ice \citep{Tsang2016,dePater2020}. If the atmosphere is primarily sublimation-supported, then during the eclipse phase of Io's orbit a portion of the \ce{SO2} atmosphere would freeze back onto the surface \citep{Saur2004}, perhaps even enough to change the near-surface bound atmosphere from collisional to non-collisional \citep{Tsang2016}.

Figure \ref{fig:column-density-timeseries} shows the best-fit model column densities during eclipse for the three species model containing O, \ce{O2} (an unknown multiple of the \ce{SO} column) and \ce{SO2}. While the O and \ce{O2} column densities exhibit some short-timescale variability, they are relatively constant over the duration of the eclipse. Further, though there are some decreases in \ce{SO2} column density with time (primarily between 30 and 75 minutes after eclipse ingress), most are followed by a proportional increase well before eclipse egress (typically around 2 hours after ingress). Though this could be explained by the collapse and recovery of some fraction of the \ce{SO2} atmosphere \citep{Tsang2016,dePater2020}, the correlation between proximity to Jupiter's limb and the relative decrease in brightness suggests this may instead be a background subtraction artifact.

\begin{figure}
    \centering
    \includegraphics[width=\columnwidth]{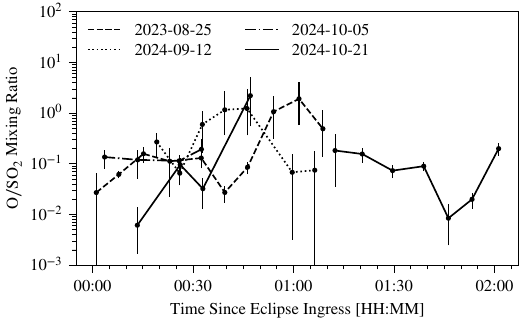}
    \caption{Best-fit emission model $\ce{O}/\ce{SO2}$ mixing ratio calculated from the column densities in Figure \ref{fig:column-density-timeseries}. Points with lower uncertainty bounds intercepting zero are cut off by the lower axis limit. Though the mixing ratio just after eclipse ingress is typically about 10\%, it can vary by more than an order of magnitude.}
    \label{fig:o-so2-timeseries}
\end{figure}

The average O column densities are on the order of $3\times 10^{18}~\mathrm{cm}^{-2}$, about a factor of 4 larger than that in the \citet{Dols2024} model and a factor of 2 smaller than the column density derived by \citet{Schmidt2023}. The maximum of the \ce{SO2} column densities are approximately $2\times 10^{19}~\mathrm{cm}^{-2}$, a factor of about 30 smaller than that in the \citet{Dols2024} model and around the lower end of typical observed column densities (see Table 19.1 and associated references in \citealt{McGrath2004} and Table 1 in \citealt{deKleer2024}). Figure \ref{fig:o-so2-timeseries} shows how the $\ce{O}/\ce{SO2}$ column mixing ratio changes over the course of the eclipses. The ratio is typically around 10\%, but varies by an order of magnitude or more and can approach parity due to the drop in \ce{SO2} column density near the eclipse midpoint.

However, the relative stability of the \ce{SO2} column and its density at the low end of previously-derived values suggests that the portion of the emission from dissociative excitation of \ce{SO2} may originate from the high altitude portion of the column, while the near-surface emission comes primarily from direct excitation of the atomic oxygen component. This would permit the near-surface \ce{SO2} atmosphere to collapse while maintaining the emitting column. This could also explain the apparent delay in the drop in \ce{SO2} emission, since the lower atmosphere would have to collapse first before the higher altitude emitting column. However, as noted above, this does not account for the apparent ``recovery'' which occurs in most of the observations after one hour in eclipse.

In such a scenario, electron energies near Io's surface may be insufficient to dissociate \ce{SO2} but are able to excite the lower-energy forbidden emissions in the atomic O, and the much higher electron number density within the ionosphere is sufficient to produce measurable emission despite quenching. The presence of the 630.0 and 636.4~nm [\OI{}] emissions at altitudes near Io's surface in the ISS images (see Section \ref{sec:cassini-comparison}) indicates that electrons are penetrating into Io's ionosphere and must therefore be losing energy as they precipitate.

\subsection{Quenching}

Laboratory experiments show \ce{SO2} quenches \ce{O(^1D)} with a rate coefficient of $(2.17\pm0.19)\times 10^{-16}~\mathrm{m^3~molecule^{-1}~s^{-1}}$ \citep{Zhao2010}. The \ce{O(^1D2 -> ^3P_2)} 630.0~nm transition has a lifetime of 178~s \citep{Wiese1996}, so the critical density at which \ce{SO2} quenches this emission (the density of \ce{SO2} for which the collisional deexcitation rate equals the auroral photon emission rate) is $2.59\times 10^{13}~\mathrm{molecules~m^{-3}}$. The \ce{O(^1D2 -> ^3P_1)} 636.4~nm transition has a much longer lifetime of 549~s \citep{Wiese1996}, so the critical density at which \ce{SO2} quenches this emission is $8.39\times 10^{12}~\mathrm{molecules~m^{-3}}$. For the atmospheric columns in the \citet{Dols2024} model, this quenching rate would limit the modeled atomic O column to $6.059 \times 10^{17}~\mathrm{m^{-2}}$ for the emission of 630.0~nm photons (83.5\% of the total column, corresponding to a minimum altitude for emission of approximately 200~km) and $5.761 \times 10^{17}~\mathrm{m^{-2}}$ for the emission of 636.4~nm photons (79.4\% of the total column, corresponding to a minimum altitude for emission of approximately 250~km).

No laboratory measurements have been published of rate coefficient for the quenching of \ce{O(^1S)} by \ce{SO2}. However, measurements have been published for \ce{O2} quenching of both \ce{O(^1D)} \citep{Streit1976} and \ce{O(^1S)} \citep{Atkinson1972,Slanger1972}. At a temperature of 115~K, \ce{O2} quenches \ce{O(^1D)} with a rate coefficient of $5.2\times 10^{-17}~\mathrm{m^3~molecule^{-1}~s^{-1}}$ and \ce{O(^1S)} with a rate coefficient of $2.9\times 10^{-21}~\mathrm{m^3~molecule^{-1}~s^{-1}}$. These rates yield critical \ce{O2} densities of $1.1\times 10^{14}~\mathrm{molecules~m^{-3}}$ for collisional deexcitation of \ce{O(^1D)} preventing the 630.0~nm transition, $3.5\times 10^{13}~\mathrm{molecules~m^{-3}}$ for collisional deexcitation of \ce{O(^1D)} preventing the 636.4~nm transition and $4.3\times 10^{20}~\mathrm{molecules~m^{-3}}$ for collisional deexcitation of \ce{O(^1S)} preventing the 557.7~nm transition. The \ce{O(^1S)} quenching rate is different by more than four orders of magnitude compared to the \ce{O(^1D)} quenching rate; assuming a similar difference in magnitude exists for \ce{SO2} there should be effectively no collisional deexcitation of \ce{O(^1S)} along the entire O column. 

The sulfur equivalent to the $630.0/557.7$~nm [\OI{}] ratio is the $1082.0/772.5$~nm [\SI{}] ratio. \citet{dePater2024} calculated a disk-integrated 1082.0~nm [\SI{}] brightness of 4.2~kR in eclipse in August 2023. When evaluated in conjunction with the 772.5~nm [\SI{}] brightness we observed with HIRES in August 2023, the ratio is consistent with emission from an atomic S column impacted by electrons with energies between 4 and 5~eV. The near-surface emissions in the ISS images (see Figure \ref{fig:cassini-filter-images}) suggest little if any quenching of the emission, though the corresponding increase in electron density makes firm conclusions difficult and altitude estimates for the exobase vary substantially \citep{Summers1996}.

\subsection{Electron Scale Height}
\label{sec:electron-scale-height}

\begin{figure}
    \centering
    \includegraphics[width=\linewidth]{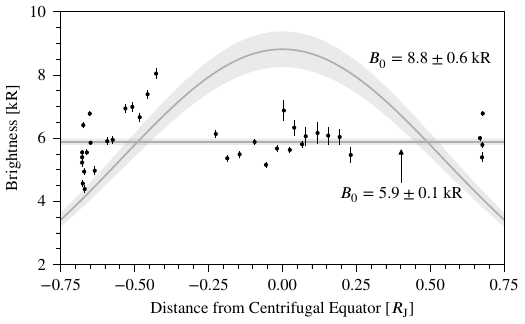}
    \caption{Relationship between Io's 630.0~nm auroral brightness and distance from the centrifugal equator. The gray lines show fits of both Equation \eqref{eq:scale-height} with the scale height fixed at 0.77~$R_\mathrm{J}$ and a simple constant model. The shaded regions show the uncertainty in the fit (for Equation \ref{eq:scale-height} this corresponds to the amplitude $n_0$, for the constant this corresponds to the uncertainty in its value). These data suggest the inherent variability of torus electron density with both longitude and on secular timescales may be on the order of factors of 2 to 3. However, the small quantity of observed brightnesses and the gaps in sampling the distance from the centrifugal equator prevent firm conclusions regarding the connection between brightness and upstream plasma density. (Note that the reported uncertainties are from the weighted fit and do not reflect the standard deviation of the data.)\\[4pt] (The data used to create this figure are available.)}
    \label{fig:scale-height-dependence}
\end{figure}

The electron number density in the Jovian magnetosphere forms a Gaussian-like vertical profile of the form
\begin{equation}\label{eq:scale-height}
    n(r, d) = n_0(r)\,\mathrm{e}^{-(d/H)^2},
\end{equation}
where $n_0$ is the number density at the centrifugal equator, $H$ is the scale height \citep[equal to $0.77~R_\mathrm{J}$ at Io's average orbital distance,][]{Bagenal2011} $r$ is Io's instantaneous orbital distance from Jupiter and $d$ is Io's distance from the centrifugal equator \citep{Hill1976}. Figure \ref{fig:scale-height-dependence} shows a fit of Equation \eqref{eq:scale-height} to the disk-integrated 630.0~nm [\OI{}] brightnesses, assuming the brightness is directly proportional to the local electron number density $n$. The six nights of observations captured Io at a variety of distances from the centrifugal equator, but some large gaps remain unobserved. Though the data between $-0.4$ and $-0.75~R_\mathrm{J}$ show some correlation between auroral brightness and electron density, the complete data across all distances are characterized effectively as well by a constant (the other fit in Figure \ref{fig:scale-height-dependence}). While this could be interpreted as suggesting there may be no intrinsic connection between the density of the upstream electrons in the torus and the brightness of the auroral emission, it could also point to variability in the density of plasma torus electrons both longitudinally within the torus and on long-term secular timescales. The latter interpretation is consistent with the results of \citet{Steffl2006}, \citet{Coffin2020} and \citet{Oliversen2001}, the latter of which saw variability in Io's sunlight 630.0~nm [\OI{}] emission by factors of 2 to 3 as a function of System~\textsc{iii} longitude (see their Figure 4). Regardless, the HIRES data do not show a strong correlation between the upstream electron plasma density and auroral brightness.

The connection between the ambient upstream plasma density and auroral brightness has been evaluated before at Io, Europa and Ganymede. \citet{Roth2014} fit UV observations of Io's 130.4 \OI{}, 135.6~nm \OI{}] and 147.9~nm \SI{} emissions and found electron scale heights between 0.8 and 1.0 $R_\mathrm{J}$, though their observations exhibited considerable variance. At Europa, \citet{Roth2016} found some evidence of an exponential decrease in the \OI{} 130.4 and \OI{}] 135.6~nm brightnesses with distance from the centrifugal equator (see their Figures 5c and 5d), but their data also showed a large variance. \citet{deKleer2023} observed a weak correlation between Europa's disk-integrated 630.0~nm aurora brightness and the plasma scale height at its orbit.

Others have shown a connection between the column densities of electrons in the flux tubes above and below a satellite at different positions within the plasma torus. \citet{Roth2016} showed a hemispheric asymmetry in the spatial distribution of the brightness which correlated with Europa's position above or below the plasma sheet centrifugal equator (see their Figure 10), suggesting either excitation from bounce motion along the flux tube or a strong gradient in plasma density within the magnetosphere between Europa's northern and southern hemispheres. \citet{Milby2024} showed that the north-south hemispheric brightness ratio on Ganymede was strongly correlated with its position within the plasma sheet, and specifically with the column density of the flux tube above/below each hemisphere using the \citet{Bagenal2011} plasma scale height at its orbital distance from Jupiter. Unfortunately, the HIRES Io data are likely dominated by emission from the equatorial glow and do not have sufficient spatial resolution to spatially-resolve Io's north-south hemispheric limb emission, so we cannot say whether an asymmetry is present in the 630.0~nm HIRES data. 

\citet{Retherford2003} attributed an observed north-south asymmetry in Io's \OI{}] 135.6~nm limb glow to the different electron column densities above and below each hemisphere, which vary with Io's position relative to the centrifugal equator. Auroral excitation would therefore come from electron bounce motion along the flux tubes above and below Io rather than impact of upstream torus electrons due to Jupiter's rotating magnetosphere. This would result in a change in the effective electron column density between the northern and southern hemispheres (and therefore a hemispheric brightness asymmetry), but should not change the total disk-integrated brightness.

Consequently, we cannot constrain the plasma scale height with the limited sampling of System \textsc{iii} magnetic latitudes available through our current set of observations. Additional observations would help to characterize both the random day-to-day variability in the brightnesses and the systematic variability with magnetic longitude.


\begin{figure}
    \centering
    \includegraphics[width=\linewidth]{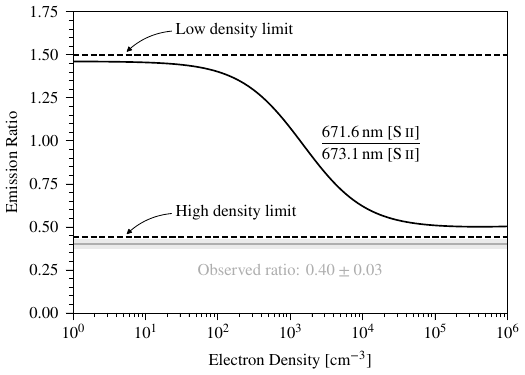}
    \caption{Dependence of the $671.6~\mathrm{nm}/673.1~\mathrm{nm}$ [\SII{}] brightness ratio on electron number density assuming excitation by 5~eV electrons. The dashed lines show the low and high-density limits defined in Equations \eqref{eq:low-density-limit} and \eqref{eq:high-density-limit}. The solid gray line and shaded region shows the observed ratio of $0.40 \pm 0.03$ which we calculated as the weighted average of the brightnesses in Table \ref{tab:all-lines}. Figure based on \citet{Osterbrock2006} Figure 5.8. Calculated using CHIANTI v10.1 \citep{Dere1997,Dere2023}.\\[4pt] (The data used to create this figure are available.)}
    \label{fig:s+-density-ratio}
\end{figure}

\subsection{Electron Number Density}
\label{sec:electron-density}

Auroral photon emission rates are proportional to both the number density of the exciting electrons and the atmospheric column density, and consequently those two quantities are degenerate. Column densities for the major components of Io's atmosphere are not well constrained and likely variable on both short and long timescales \citep[see Table 19.1 in][]{McGrath2004}, limiting the ability for our aurora model to accurately determine the required electron density necessary to produce the brightnesses we observe. Further, our model assumes only direct excitation of neutral atoms and dissociative excitation of neutral molecules and does not currently incorporate other processes which can lead to auroral emission like dissociative recombination of molecular ions.

However, the ratio between certain ion emission lines can be used to derive electron density without the need to constrain the atmospheric column if those emission lines share the same ground level and have upper levels very close in energy. Two such lines frequently used to study Io's plasma torus are the forbidden 671.6~nm \ce{S+}(\ce{^2D^$\mathrm{o}$_${5/2}$ -> ^4S^$\mathrm{o}$_${3/2}$}) and 673.1~nm \ce{S+}(\ce{^2D^$\mathrm{o}$_${3/2}$ -> ^4S^$\mathrm{o}$_${3/2}$}) emissions \citep[e.g.,][]{Brown1982,Kuppers1997,Kuppers2000,Nerney2024}. Figure \ref{fig:s+-density-ratio} shows how the ratio between these two emissions changes with electron densities between 1 and $10^6~\mathrm{cm}^{-3}$. At low electron densities (below around 100~$\mathrm{cm^{-3}}$) the relative populations of the upper states are determined by their statistical weights $2J+1$, so the emission ratio limits to
\begin{equation}\label{eq:low-density-limit}
    \frac{2J_{671.6~\mathrm{nm}}+1}{2J_{673.1~\mathrm{nm}}+1} = \frac{2(5/2)+1}{2(3/2)+1} = 1.5.
\end{equation}
At larger electron densities, electron collisions begin to deexcite the ions before they have a change to radiate. Because the 671.6~nm state has a longer lifetime than the 673.1~nm state, collisions preferentially deexcite those ions, and the emission ratio begins to decrease. Above an electron density of approximately $10^4~\mathrm{cm^{-3}}$, collisional deexcitation dominates for both states and emission therefore comes only from the few atoms that survive long enough to radiate. In this case, the emission ratio is modulated by the relative lifetimes $\tau$ of the states, equal to the inverse of the Einstein A coefficients for those emissions ($\tau = 1/A$):
\begin{equation}\label{eq:high-density-limit}
    \frac{\left(2J_{671.6~\mathrm{nm}}+1\right)/\tau_{671.6~\mathrm{nm}}}{\left(2J_{673.1~\mathrm{nm}}+1\right)/\tau_{673.1~\mathrm{nm}}} =1.5\left(\frac{2.02\times 10^{-4}~\mathrm{s}^{-1}}{6.84\times 10^{-4}~\mathrm{s}^{-1}}\right) = 0.44.
\end{equation}
For the 671.6~nm transition, $A_{671.6~\mathrm{nm}}$ is the sum of the Einstein A coefficient for electric quadrupole emission $1.88\times 10^{-4}~\mathrm{s}^{-1}$ and magnetic dipole emission $1.39\times 10^{-5}~\mathrm{s}^{-1}$ \citep{Podobedova2009}. Similarly, for the 673.1~nm transition, $A_{673.1~\mathrm{nm}}$ is the sum of the Einstein A coefficient for electric quadrupole emission $1.21\times 10^{-4}~\mathrm{s}^{-1}$ and magnetic dipole emission $5.63\times 10^{-5}~\mathrm{s}^{-1}$ \citep{Podobedova2009}. Dashed gray lines in Figure \ref{fig:s+-density-ratio} show these limits; the ratio (black line) doesn't exactly reach these limits due to additional processes modeled by CHIANTI which produce excited \ce{S+} like recombination of \ce{S^2+}.

The region where the emission ratio is sensitive to changing electron densities (between 100 and $10^4~\mathrm{cm^{-3}}$) includes the range of electron densities expected in the upstream plasma torus \citep[1200 to 3800~$\mathrm{cm^{-3}}$,][]{Bagenal2020}. However, the observed ratios (see Table \ref{tab:all-lines} and Figure \ref{fig:s+-density-ratio}) are all below the high-density limit at any electron energy. Io's Alfv\'{e}n wings have much higher electron number densities (between $1.53\times 10^{4}$ and $3.1\times 10^{4}~\mathrm{cm}^{-3}$) compared to the surrounding plasma torus \citep{Buccino2025}. However, the observed ratio from our data is still well below the predicted ratio even for their higher electron number density estimates.

These results indicate there must be either collisional quenching or additional sources of excited \ce{S+} beyond electron impact on an existing ion column like charge exchange or dissociative recombination of \ce{SO+} and/or \ce{SO2+}. We also looked at the modeled ratios of the other \ce{S+} emissions at 406.9 and 407.6~nm and the \ce{O+} emissions at 732.0 and 733.0~nm, but neither exhibited notable variability across the range of electron energies and densities plausible for Io's ionosphere and extended atmosphere and were therefore not useful diagnostics of either electron energy or density.

\section{Conclusions}
In this study, we analyzed visible and near-infrared auroral emissions from Io taken while it was in eclipse by Jupiter. From these spectra we isolated optical wavelength emissions from a variety of neutral and singly-ionized atoms never before spectroscopically detected at Io. These included the \OI{} lines at 777.4 and 844.6~nm, the \NaI{} lines at 818.3 and 819.5~nm, the [\SI{}] lines at 458.9 and 772.5~nm, the \SI{} triplet at 922.3~nm, the [\OII{}] lines at 732.0 and 733.0~nm and the [\SII{}] lines at 406.9, 407.6, 671.6 and 673.1~nm. These new detections more than tripled the number of identified emission lines from Io at visible and near-infrared wavelengths.

We compared our spectral data with Cassini/ISS images of Io in eclipse to interpret the spatial distribution of the various emissions. We determined that the limb glow was consistent with electron impact on the atomic coronae, while the equatorial spots were likely dominated by emission from dissociative impact on the molecular atmosphere. Volcanic plumes exhibited emissions from Na, K, \ce{O^+} and \ce{S^+}.

The detection of additional optical emission lines allowed us to use our auroral emission model to determine the atmospheric species producing the auroral oxygen emissions. We found a three-species atmosphere composed of O, \ce{SO2} and \ce{O2} (which we interpreted as a proxy for \ce{SO}) reproduced the observed emission brightnesses. The derived column densities for \ce{SO2} were consistent with previous studies, but they did not show a clear, systematic decrease after eclipse ingress, suggesting the molecular atmospheric columns producing auroral emission may be restricted to higher altitudes than the atomic columns. This could be interpreted as evidence for collisional quenching of O at lower altitudes, though the Cassini/ISS images isolating forbidden emissions with long lifetimes show emission all the way down to the surface. Alternatively, this could suggest an excitation process more complex than just electron impact on full neutral atomic and molecular columns. In particular, the fraction of the emission resulting from \ce{SO2} may occur at high altitudes where the electron energy is still large enough to produce dissociative excitation, while the low-altitude emission comes from electron impact on atomic O excited by the lower-energy electrons within Io's ionosphere.

We compared the connection between [\OI{}] 630.0~nm auroral emission and ambient magnetospheric electron density, but found ambiguous evidence for a direct correlation. This could be explained by some combination of System~\textsc{iii} longitudinal variability in the plasma density or long-term secular variation in the atmospheric column density and/or electron number density. Alternatively, it could also indicate that the excitation came from electron bounce motion within Io's flux tube and therefore the disk-integrated brightness (and particularly the equatorial glow) were excited by the full electron column rather than the local electron number density. If so, the brightness would be insensitive to Io's vertical position within the torus. The small number of observations currently available prevented us from determining which effect dominates Io's auroral emission. A more comprehensive set of observations covering the full range of System~\textsc{iii} magnetic longitudes would help determine if there is additional longitudinal variability in density.


Finally, we analyzed \ce{S+} emission line ratios in order to determine the number density of the emission-exciting electrons, but found the observed ratios were incompatible with electron impact on a pure \ce{S+} column, instead requiring additional sources of excited \ce{S+} such as dissociation of ionospheric \ce{SO+} and \ce{SO2+}. Consequently, the number density cannot be constrained from these data.

\section*{Acknowledgments}
This work benefited from scientific exchanges that took place within International Space Sciences Institute (ISSI) international Team \#559 and as part of an ISSI workshop, Team \#515.

Z.M. was supported by the NASA Future Investigators in NASA Earth and Space Science and Technology (FINESST) program grant number 80NSSC24K1721 and through program JWST-GO-04078.001-A provided by NASA through a grant from the Space Telescope Science Institute, which is operated by the Association of Universities for Research in Astronomy, Inc., under NASA contract NAS 5-03127. C.S. acknowledges support from NASA programs 80NSSC22K0954, 80NSSC21K1138 and from the NSF under program AST-2108416.

The authors thank Dr. Kyle Connour for his insightful and constructive comments on the manuscript.

The data presented herein were obtained at the W.~M. Keck Observatory, which is operated as a scientific partnership among the California Institute of Technology, the University of California and the National Aeronautics and Space Administration. The Observatory was made possible by the generous financial support of the W.~M. Keck Foundation.

This research has made use of the Keck Observatory Archive (KOA), which is operated by the W.~M. Keck Observatory and the NASA Exoplanet Science Institute (NExScI), under contract with the National Aeronautics and Space Administration.

The authors wish to recognize and acknowledge the very significant cultural role and reverence that the summit of Maunakea has always had within the indigenous Hawaiian community. We are most fortunate to have the opportunity to conduct observations from this mountain.

\facilities{Keck:I}

\bibliography{references}{}
\bibliographystyle{aasjournal}

\appendix

\section{Cassini/ISS Image Files}
Table \ref{tab:app-cassini-image-filenames} lists the filenames of Cassini/ISS Narrow Angle Camera images displayed in Figures \ref{fig:cassini-morphology} and \ref{fig:cassini-filter-images}. We accessed these images through the Planetary Data System Image Atlas at \url{https://pds-imaging.jpl.nasa.gov/search}.
\begin{table}[H]
    \centering
    \caption{File names of Cassini/ISS Images}
    \label{tab:app-cassini-image-filenames}
    \begin{tabular}{lll}
        \toprule
        Filter & UTC Time\textsuperscript{\scriptsize{a}} & File Name \\
        \midrule
        CLR & 12:44:23 & \texttt{N1357390564\_1} \\
        BL1 & 11:41:15 & \texttt{N1357386776\_1} \\
        GRN & 11:37:02 & \texttt{N1357386523\_1} \\
        RED+GRN & 11:37:51 & \texttt{N1357386572\_1} \\
        CB1 & 11:39:29 & \texttt{N1357386670\_1} \\
        RED & 11:42:04 & \texttt{N1357386825\_1} \\
        RED+IR1 & 11:38:40 & \texttt{N1357386621\_1} \\
        IR1 & 11:43:00 & \texttt{N1357386881\_1} \\
        IR2 & 11:43:56 & \texttt{N1357386937\_1} \\
        IR3 & 11:44:52 & \texttt{N1357386993\_1} \\
        \bottomrule
        \multicolumn{3}{p{0.75\linewidth}}{
        \vspace*{-5pt}
        \begin{itemize}[nosep, labelsep=-1.5pt, align=left, leftmargin=*]
        \item[\textsuperscript{a}]At the midpoint of the observation on 2001 January 05.
        \end{itemize}}
    \end{tabular}
\end{table}

\section{HIRES Data Files}
Tables \ref{tab:files-2022-11-24} through \ref{tab:files-2024-10-21} lists the file names and corresponding observation type for each FITS file used in this study. All data are available from the Keck Observatory Archive (KOA)\footnote{\url{http://koa.ipac.caltech.edu}}. These data were taken as a part of the program ``Joint Keck-Juno observations of Jupiter, its moons and its magnetosphere'' with program IDs N059 and N018 under principal investigator Carl Schmidt (the 2022 and 2023 observations) and N078 under principal investigator Luke Moore (the 2024 observations).

\LTcapwidth=\columnwidth
\begin{longtable}{lll}
\caption{Data files from 2022 November 24 used in this study and their corresponding observation type and target. All Io observations were taken during eclipse.}\label{tab:files-2022-11-24}
\endfirsthead
\toprule
KOA Unique File Name \relax& Type \relax& Target\\
\midrule
\texttt{HI.20221124.12738.11.fits.gz} \relax& Calibration \relax& None (bias)\\
\texttt{HI.20221124.12782.99.fits.gz} \relax& Calibration \relax& None (bias)\\
\texttt{HI.20221124.12827.36.fits.gz} \relax& Calibration \relax& None (bias)\\
\texttt{HI.20221124.12871.73.fits.gz} \relax& Calibration \relax& None (bias)\\
\texttt{HI.20221124.12916.61.fits.gz} \relax& Calibration \relax& None (bias)\\
\texttt{HI.20221124.12960.98.fits.gz} \relax& Calibration \relax& None (bias)\\
\texttt{HI.20221124.13005.35.fits.gz} \relax& Calibration \relax& None (bias)\\
\texttt{HI.20221124.13050.23.fits.gz} \relax& Calibration \relax& None (bias)\\
\texttt{HI.20221124.13094.90.fits.gz} \relax& Calibration \relax& None (bias)\\
\texttt{HI.20221124.13138.97.fits.gz} \relax& Calibration \relax& None (bias)\\
\texttt{HI.20221124.13209.35.fits.gz} \relax& Calibration \relax& Quartz flat lamp\\
\texttt{HI.20221124.13254.87.fits.gz} \relax& Calibration \relax& Quartz flat lamp\\
\texttt{HI.20221124.13300.13.fits.gz} \relax& Calibration \relax& Quartz flat lamp\\
\texttt{HI.20221124.13346.54.fits.gz} \relax& Calibration \relax& Quartz flat lamp\\
\texttt{HI.20221124.13391.93.fits.gz} \relax& Calibration \relax& Quartz flat lamp\\
\texttt{HI.20221124.13437.32.fits.gz} \relax& Calibration \relax& Quartz flat lamp\\
\texttt{HI.20221124.13482.71.fits.gz} \relax& Calibration \relax& Quartz flat lamp\\
\texttt{HI.20221124.13528.10.fits.gz} \relax& Calibration \relax& Quartz flat lamp\\
\texttt{HI.20221124.13573.49.fits.gz} \relax& Calibration \relax& Quartz flat lamp\\
\texttt{HI.20221124.13618.88.fits.gz} \relax& Calibration \relax& Quartz flat lamp\\
\texttt{HI.20221124.13688.75.fits.gz} \relax& Calibration \relax& ThAr arc lamp\\
\texttt{HI.20221124.13734.87.fits.gz} \relax& Calibration \relax& ThAr arc lamp\\
\texttt{HI.20221124.13779.26.fits.gz} \relax& Calibration \relax& ThAr arc lamp\\
\texttt{HI.20221124.13824.65.fits.gz} \relax& Calibration \relax& ThAr arc lamp\\
\texttt{HI.20221124.13870.40.fits.gz} \relax& Calibration \relax& ThAr arc lamp\\
\texttt{HI.20221124.13915.43.fits.gz} \relax& Calibration \relax& ThAr arc lamp\\
\texttt{HI.20221124.13960.82.fits.gz} \relax& Calibration \relax& ThAr arc lamp\\
\texttt{HI.20221124.14006.21.fits.gz} \relax& Calibration \relax& ThAr arc lamp\\
\texttt{HI.20221124.14051.60.fits.gz} \relax& Calibration \relax& ThAr arc lamp\\
\texttt{HI.20221124.14096.99.fits.gz} \relax& Calibration \relax& ThAr arc lamp\\
\midrule
\texttt{HI.20221124.15491.84.fits.gz} \relax& Science \relax& HD 218639\\
\texttt{HI.20221124.15777.44.fits.gz} \relax& Science \relax& Jupiter\\
\texttt{HI.20221124.17003.48.fits.gz} \relax& Science \relax& Io\\
\texttt{HI.20221124.17417.90.fits.gz} \relax& Science \relax& Ganymede\\
\texttt{HI.20221124.17490.20.fits.gz} \relax& Science \relax& Io\\
\texttt{HI.20221124.17903.12.fits.gz} \relax& Science \relax& Ganymede\\
\texttt{HI.20221124.18032.66.fits.gz} \relax& Science \relax& Io\\
\texttt{HI.20221124.18394.25.fits.gz} \relax& Science \relax& Ganymede\\
\texttt{HI.20221124.18590.60.fits.gz} \relax& Science \relax& Io\\
\texttt{HI.20221124.18944.54.fits.gz} \relax& Science \relax& Ganymede\\
\texttt{HI.20221124.19033.28.fits.gz} \relax& Science \relax& Io\\
\texttt{HI.20221124.19389.77.fits.gz} \relax& Science \relax& Ganymede\\
\texttt{HI.20221124.19457.90.fits.gz} \relax& Science \relax& Io\\
\texttt{HI.20221124.19839.80.fits.gz} \relax& Science \relax& Ganymede\\
\texttt{HI.20221124.19932.92.fits.gz} \relax& Science \relax& Io\\
\texttt{HI.20221124.20279.21.fits.gz} \relax& Science \relax& Ganymede\\
\bottomrule
\end{longtable}

\LTcapwidth=\columnwidth
\begin{longtable}{lll}
\caption{Data files from 2023 August 09 used in this study and their corresponding observation type and target. All Io observations were taken during eclipse.}\label{tab:files-2023-08-09}
\endfirsthead
\toprule
KOA Unique File Name \relax& Type \relax& Target\\
\midrule
\texttt{HI.20230809.07390.58.fits.gz} \relax& Calibration \relax& None (bias)\\
\texttt{HI.20230809.07434.95.fits.gz} \relax& Calibration \relax& None (bias)\\
\texttt{HI.20230809.07479.32.fits.gz} \relax& Calibration \relax& None (bias)\\
\texttt{HI.20230809.07523.69.fits.gz} \relax& Calibration \relax& None (bias)\\
\texttt{HI.20230809.07569.80.fits.gz} \relax& Calibration \relax& None (bias)\\
\texttt{HI.20230809.07613.45.fits.gz} \relax& Calibration \relax& None (bias)\\
\texttt{HI.20230809.07658.33.fits.gz} \relax& Calibration \relax& None (bias)\\
\texttt{HI.20230809.07702.70.fits.gz} \relax& Calibration \relax& None (bias)\\
\texttt{HI.20230809.07747.70.fits.gz} \relax& Calibration \relax& None (bias)\\
\texttt{HI.20230809.07791.44.fits.gz} \relax& Calibration \relax& None (bias)\\
\texttt{HI.20230809.08340.20.fits.gz} \relax& Calibration \relax& ThAr arc lamp\\
\texttt{HI.20230809.08385.59.fits.gz} \relax& Calibration \relax& ThAr arc lamp\\
\texttt{HI.20230809.08430.98.fits.gz} \relax& Calibration \relax& ThAr arc lamp\\
\texttt{HI.20230809.08475.86.fits.gz} \relax& Calibration \relax& ThAr arc lamp\\
\texttt{HI.20230809.08521.76.fits.gz} \relax& Calibration \relax& ThAr arc lamp\\
\texttt{HI.20230809.08567.66.fits.gz} \relax& Calibration \relax& ThAr arc lamp\\
\texttt{HI.20230809.08613.50.fits.gz} \relax& Calibration \relax& ThAr arc lamp\\
\texttt{HI.20230809.08659.46.fits.gz} \relax& Calibration \relax& ThAr arc lamp\\
\texttt{HI.20230809.08705.36.fits.gz} \relax& Calibration \relax& ThAr arc lamp\\
\texttt{HI.20230809.08751.26.fits.gz} \relax& Calibration \relax& ThAr arc lamp\\
\texttt{HI.20230809.09590.22.fits.gz} \relax& Calibration \relax& Quartz flat lamp\\
\texttt{HI.20230809.09638.16.fits.gz} \relax& Calibration \relax& Quartz flat lamp\\
\texttt{HI.20230809.09684.57.fits.gz} \relax& Calibration \relax& Quartz flat lamp\\
\texttt{HI.20230809.09732.00.fits.gz} \relax& Calibration \relax& Quartz flat lamp\\
\texttt{HI.20230809.09778.92.fits.gz} \relax& Calibration \relax& Quartz flat lamp\\
\texttt{HI.20230809.09825.33.fits.gz} \relax& Calibration \relax& Quartz flat lamp\\
\texttt{HI.20230809.09872.77.fits.gz} \relax& Calibration \relax& Quartz flat lamp\\
\texttt{HI.20230809.09919.69.fits.gz} \relax& Calibration \relax& Quartz flat lamp\\
\texttt{HI.20230809.09966.61.fits.gz} \relax& Calibration \relax& Quartz flat lamp\\
\texttt{HI.20230809.10013.20.fits.gz} \relax& Calibration \relax& Quartz flat lamp\\
\midrule
\texttt{HI.20230809.47295.64.fits.gz} \relax& Science \relax& HD 13869\\
\texttt{HI.20230809.47859.19.fits.gz} \relax& Science \relax& Jupiter\\
\texttt{HI.20230809.49839.52.fits.gz} \relax& Science \relax& Io\\
\texttt{HI.20230809.50210.80.fits.gz} \relax& Science \relax& Europa\\
\texttt{HI.20230809.50269.45.fits.gz} \relax& Science \relax& Io\\
\texttt{HI.20230809.50622.37.fits.gz} \relax& Science \relax& Europa\\
\texttt{HI.20230809.50708.50.fits.gz} \relax& Science \relax& Io\\
\texttt{HI.20230809.51054.85.fits.gz} \relax& Science \relax& Europa\\
\texttt{HI.20230809.51108.91.fits.gz} \relax& Science \relax& Io\\
\texttt{HI.20230809.51478.15.fits.gz} \relax& Science \relax& Europa\\
\texttt{HI.20230809.51548.20.fits.gz} \relax& Science \relax& Io\\
\texttt{HI.20230809.51895.33.fits.gz} \relax& Science \relax& Europa\\
\texttt{HI.20230809.51977.44.fits.gz} \relax& Science \relax& Io\\
\texttt{HI.20230809.52324.75.fits.gz} \relax& Science \relax& Europa\\
\texttt{HI.20230809.52379.83.fits.gz} \relax& Science \relax& Io\\
\texttt{HI.20230809.52749.58.fits.gz} \relax& Science \relax& Europa\\
\bottomrule
\end{longtable}

\LTcapwidth=\columnwidth
\begin{longtable}{lll}
\caption{Data files from 2023 August 25 used in this study and their corresponding observation type and target. All Io observations were taken during eclipse.}\label{tab:files-2023-08-25}
\endfirsthead
\toprule
KOA Unique File Name \relax& Type \relax& Target\\
\midrule
\texttt{HI.20230825.11285.49.fits.gz} \relax& Calibration \relax& None (bias)\\
\texttt{HI.20230825.11329.35.fits.gz} \relax& Calibration \relax& None (bias)\\
\texttt{HI.20230825.11373.21.fits.gz} \relax& Calibration \relax& None (bias)\\
\texttt{HI.20230825.11418.90.fits.gz} \relax& Calibration \relax& None (bias)\\
\texttt{HI.20230825.11462.46.fits.gz} \relax& Calibration \relax& None (bias)\\
\texttt{HI.20230825.11506.83.fits.gz} \relax& Calibration \relax& None (bias)\\
\texttt{HI.20230825.11551.71.fits.gz} \relax& Calibration \relax& None (bias)\\
\texttt{HI.20230825.11596.80.fits.gz} \relax& Calibration \relax& None (bias)\\
\texttt{HI.20230825.11640.45.fits.gz} \relax& Calibration \relax& None (bias)\\
\texttt{HI.20230825.11684.82.fits.gz} \relax& Calibration \relax& None (bias)\\
\texttt{HI.20230825.11755.20.fits.gz} \relax& Calibration \relax& Quartz flat lamp\\
\texttt{HI.20230825.11801.10.fits.gz} \relax& Calibration \relax& Quartz flat lamp\\
\texttt{HI.20230825.11846.49.fits.gz} \relax& Calibration \relax& Quartz flat lamp\\
\texttt{HI.20230825.11892.39.fits.gz} \relax& Calibration \relax& Quartz flat lamp\\
\texttt{HI.20230825.11938.29.fits.gz} \relax& Calibration \relax& Quartz flat lamp\\
\texttt{HI.20230825.11983.68.fits.gz} \relax& Calibration \relax& Quartz flat lamp\\
\texttt{HI.20230825.12030.90.fits.gz} \relax& Calibration \relax& Quartz flat lamp\\
\texttt{HI.20230825.12075.48.fits.gz} \relax& Calibration \relax& Quartz flat lamp\\
\texttt{HI.20230825.12121.38.fits.gz} \relax& Calibration \relax& Quartz flat lamp\\
\texttt{HI.20230825.12167.28.fits.gz} \relax& Calibration \relax& Quartz flat lamp\\
\texttt{HI.20230825.12236.13.fits.gz} \relax& Calibration \relax& ThAr arc lamp\\
\texttt{HI.20230825.12281.52.fits.gz} \relax& Calibration \relax& ThAr arc lamp\\
\texttt{HI.20230825.12326.91.fits.gz} \relax& Calibration \relax& ThAr arc lamp\\
\texttt{HI.20230825.12372.30.fits.gz} \relax& Calibration \relax& ThAr arc lamp\\
\texttt{HI.20230825.12417.69.fits.gz} \relax& Calibration \relax& ThAr arc lamp\\
\texttt{HI.20230825.12463.80.fits.gz} \relax& Calibration \relax& ThAr arc lamp\\
\texttt{HI.20230825.12508.47.fits.gz} \relax& Calibration \relax& ThAr arc lamp\\
\texttt{HI.20230825.12553.86.fits.gz} \relax& Calibration \relax& ThAr arc lamp\\
\texttt{HI.20230825.12599.76.fits.gz} \relax& Calibration \relax& ThAr arc lamp\\
\texttt{HI.20230825.12645.15.fits.gz} \relax& Calibration \relax& ThAr arc lamp\\
\midrule
\texttt{HI.20230825.39325.85.fits.gz} \relax& Science \relax& HD 13869\\
\texttt{HI.20230825.42546.50.fits.gz} \relax& Science \relax& Io\\
\texttt{HI.20230825.42893.83.fits.gz} \relax& Science \relax& Europa\\
\texttt{HI.20230825.42950.45.fits.gz} \relax& Science \relax& Io\\
\texttt{HI.20230825.43308.98.fits.gz} \relax& Science \relax& Europa\\
\texttt{HI.20230825.43391.60.fits.gz} \relax& Science \relax& Io\\
\texttt{HI.20230825.43736.87.fits.gz} \relax& Science \relax& Europa\\
\texttt{HI.20230825.43863.35.fits.gz} \relax& Science \relax& Io\\
\texttt{HI.20230825.44221.37.fits.gz} \relax& Science \relax& Europa\\
\texttt{HI.20230825.44426.39.fits.gz} \relax& Science \relax& Io\\
\texttt{HI.20230825.44783.90.fits.gz} \relax& Science \relax& Europa\\
\texttt{HI.20230825.44847.65.fits.gz} \relax& Science \relax& Io\\
\texttt{HI.20230825.45199.61.fits.gz} \relax& Science \relax& Europa\\
\texttt{HI.20230825.45256.67.fits.gz} \relax& Science \relax& Io\\
\texttt{HI.20230825.45616.73.fits.gz} \relax& Science \relax& Europa\\
\texttt{HI.20230825.45722.30.fits.gz} \relax& Science \relax& Io\\
\texttt{HI.20230825.46124.18.fits.gz} \relax& Science \relax& Europa\\
\texttt{HI.20230825.46177.22.fits.gz} \relax& Science \relax& Io\\
\texttt{HI.20230825.46520.96.fits.gz} \relax& Science \relax& Europa\\
\texttt{HI.20230825.46605.62.fits.gz} \relax& Science \relax& Io\\
\texttt{HI.20230825.46968.74.fits.gz} \relax& Science \relax& Europa\\
\texttt{HI.20230825.55149.15.fits.gz} \relax& Science \relax& Jupiter\\
\bottomrule
\end{longtable}

\LTcapwidth=\columnwidth
\begin{longtable}{lll}
\caption{Data files from 2024 September 12 used in this study and their corresponding observation type and target. All Io observations were taken during eclipse.}\label{tab:files-2024-09-12}
\endfirsthead
\toprule
KOA Unique File Name \relax& Type \relax& Target\\
\midrule
\texttt{HI.20240912.03154.28.fits.gz} \relax& Calibration \relax& None (bias)\\
\texttt{HI.20240912.03198.65.fits.gz} \relax& Calibration \relax& None (bias)\\
\texttt{HI.20240912.03243.53.fits.gz} \relax& Calibration \relax& None (bias)\\
\texttt{HI.20240912.03287.90.fits.gz} \relax& Calibration \relax& None (bias)\\
\texttt{HI.20240912.03332.27.fits.gz} \relax& Calibration \relax& None (bias)\\
\texttt{HI.20240912.03377.15.fits.gz} \relax& Calibration \relax& None (bias)\\
\texttt{HI.20240912.03421.52.fits.gz} \relax& Calibration \relax& None (bias)\\
\texttt{HI.20240912.03465.89.fits.gz} \relax& Calibration \relax& None (bias)\\
\texttt{HI.20240912.03510.26.fits.gz} \relax& Calibration \relax& None (bias)\\
\texttt{HI.20240912.03555.14.fits.gz} \relax& Calibration \relax& None (bias)\\
\texttt{HI.20240912.03624.50.fits.gz} \relax& Calibration \relax& Quartz flat lamp\\
\texttt{HI.20240912.03669.89.fits.gz} \relax& Calibration \relax& Quartz flat lamp\\
\texttt{HI.20240912.03715.79.fits.gz} \relax& Calibration \relax& Quartz flat lamp\\
\texttt{HI.20240912.03761.69.fits.gz} \relax& Calibration \relax& Quartz flat lamp\\
\texttt{HI.20240912.03807.59.fits.gz} \relax& Calibration \relax& Quartz flat lamp\\
\texttt{HI.20240912.03852.98.fits.gz} \relax& Calibration \relax& Quartz flat lamp\\
\texttt{HI.20240912.03897.86.fits.gz} \relax& Calibration \relax& Quartz flat lamp\\
\texttt{HI.20240912.03943.76.fits.gz} \relax& Calibration \relax& Quartz flat lamp\\
\texttt{HI.20240912.03989.15.fits.gz} \relax& Calibration \relax& Quartz flat lamp\\
\texttt{HI.20240912.04035.50.fits.gz} \relax& Calibration \relax& Quartz flat lamp\\
\texttt{HI.20240912.04105.43.fits.gz} \relax& Calibration \relax& ThAr arc lamp\\
\texttt{HI.20240912.04151.33.fits.gz} \relax& Calibration \relax& ThAr arc lamp\\
\texttt{HI.20240912.04197.23.fits.gz} \relax& Calibration \relax& ThAr arc lamp\\
\texttt{HI.20240912.04242.11.fits.gz} \relax& Calibration \relax& ThAr arc lamp\\
\texttt{HI.20240912.04288.52.fits.gz} \relax& Calibration \relax& ThAr arc lamp\\
\texttt{HI.20240912.04333.91.fits.gz} \relax& Calibration \relax& ThAr arc lamp\\
\texttt{HI.20240912.04380.32.fits.gz} \relax& Calibration \relax& ThAr arc lamp\\
\texttt{HI.20240912.04425.71.fits.gz} \relax& Calibration \relax& ThAr arc lamp\\
\texttt{HI.20240912.04471.10.fits.gz} \relax& Calibration \relax& ThAr arc lamp\\
\texttt{HI.20240912.04516.49.fits.gz} \relax& Calibration \relax& ThAr arc lamp\\
\midrule
\texttt{HI.20240912.46822.57.fits.gz} \relax& Science \relax& HD 34203\\
\texttt{HI.20240912.49928.98.fits.gz} \relax& Science \relax& Io\\
\texttt{HI.20240912.50277.82.fits.gz} \relax& Science \relax& Europa\\
\texttt{HI.20240912.50341.60.fits.gz} \relax& Science \relax& Io\\
\texttt{HI.20240912.50691.94.fits.gz} \relax& Science \relax& Europa\\
\texttt{HI.20240912.50746.00.fits.gz} \relax& Science \relax& Io\\
\texttt{HI.20240912.51096.37.fits.gz} \relax& Science \relax& Europa\\
\texttt{HI.20240912.51147.37.fits.gz} \relax& Science \relax& Io\\
\texttt{HI.20240912.51493.15.fits.gz} \relax& Science \relax& Europa\\
\texttt{HI.20240912.51546.19.fits.gz} \relax& Science \relax& Io\\
\texttt{HI.20240912.52298.95.fits.gz} \relax& Science \relax& Europa\\
\texttt{HI.20240912.52356.58.fits.gz} \relax& Science \relax& Io\\
\texttt{HI.20240912.52707.46.fits.gz} \relax& Science \relax& Europa\\
\texttt{HI.20240912.52758.97.fits.gz} \relax& Science \relax& Io\\
\texttt{HI.20240912.53106.28.fits.gz} \relax& Science \relax& Europa\\
\texttt{HI.20240912.53643.31.fits.gz} \relax& Science \relax& Jupiter\\
\bottomrule
\end{longtable}

\LTcapwidth=\columnwidth
\begin{longtable}{lll}
\caption{Data files from 2024 October 05 used in this study and their corresponding observation type and target. All Io observations were taken during eclipse.}\label{tab:files-2024-10-05}
\endfirsthead
\toprule
KOA Unique File Name \relax& Type \relax& Target\\
\midrule
\texttt{HI.20241005.49655.68.fits.gz} \relax& Science \relax& Io\\
\texttt{HI.20241005.49895.38.fits.gz} \relax& Science \relax& Europa\\
\texttt{HI.20241005.50247.28.fits.gz} \relax& Science \relax& Io\\
\texttt{HI.20241005.50482.90.fits.gz} \relax& Science \relax& Europa\\
\texttt{HI.20241005.50533.39.fits.gz} \relax& Science \relax& Io\\
\texttt{HI.20241005.50892.43.fits.gz} \relax& Science \relax& Europa\\
\texttt{HI.20241005.50991.37.fits.gz} \relax& Science \relax& Io\\
\texttt{HI.20241005.51336.64.fits.gz} \relax& Science \relax& Europa\\
\texttt{HI.20241005.51388.66.fits.gz} \relax& Science \relax& Io\\
\texttt{HI.20241005.51734.45.fits.gz} \relax& Science \relax& Europa\\
\texttt{HI.20241005.51791.60.fits.gz} \relax& Science \relax& Io\\
\texttt{HI.20241005.52135.31.fits.gz} \relax& Science \relax& Europa\\
\texttt{HI.20241005.52214.36.fits.gz} \relax& Science \relax& Io\\
\texttt{HI.20241005.52562.18.fits.gz} \relax& Science \relax& Europa\\
\texttt{HI.20241005.53643.89.fits.gz} \relax& Science \relax& Jupiter\\
\texttt{HI.20241005.56670.23.fits.gz} \relax& Science \relax& HD 21686\\
\midrule
\texttt{HI.20241005.56811.50.fits.gz} \relax& Calibration \relax& None (bias)\\
\texttt{HI.20241005.56855.36.fits.gz} \relax& Calibration \relax& None (bias)\\
\texttt{HI.20241005.56900.24.fits.gz} \relax& Calibration \relax& None (bias)\\
\texttt{HI.20241005.56944.61.fits.gz} \relax& Calibration \relax& None (bias)\\
\texttt{HI.20241005.56989.49.fits.gz} \relax& Calibration \relax& None (bias)\\
\texttt{HI.20241005.57033.86.fits.gz} \relax& Calibration \relax& None (bias)\\
\texttt{HI.20241005.57078.23.fits.gz} \relax& Calibration \relax& None (bias)\\
\texttt{HI.20241005.57122.60.fits.gz} \relax& Calibration \relax& None (bias)\\
\texttt{HI.20241005.57167.48.fits.gz} \relax& Calibration \relax& None (bias)\\
\texttt{HI.20241005.57211.85.fits.gz} \relax& Calibration \relax& None (bias)\\
\texttt{HI.20241005.57319.97.fits.gz} \relax& Calibration \relax& Quartz flat lamp\\
\texttt{HI.20241005.57365.87.fits.gz} \relax& Calibration \relax& Quartz flat lamp\\
\texttt{HI.20241005.57411.77.fits.gz} \relax& Calibration \relax& Quartz flat lamp\\
\texttt{HI.20241005.57458.69.fits.gz} \relax& Calibration \relax& Quartz flat lamp\\
\texttt{HI.20241005.57504.80.fits.gz} \relax& Calibration \relax& Quartz flat lamp\\
\texttt{HI.20241005.57550.49.fits.gz} \relax& Calibration \relax& Quartz flat lamp\\
\texttt{HI.20241005.57596.39.fits.gz} \relax& Calibration \relax& Quartz flat lamp\\
\texttt{HI.20241005.57642.80.fits.gz} \relax& Calibration \relax& Quartz flat lamp\\
\texttt{HI.20241005.57689.21.fits.gz} \relax& Calibration \relax& Quartz flat lamp\\
\texttt{HI.20241005.57735.11.fits.gz} \relax& Calibration \relax& Quartz flat lamp\\
\texttt{HI.20241005.57804.98.fits.gz} \relax& Calibration \relax& ThAr arc lamp\\
\texttt{HI.20241005.57850.88.fits.gz} \relax& Calibration \relax& ThAr arc lamp\\
\texttt{HI.20241005.57896.78.fits.gz} \relax& Calibration \relax& ThAr arc lamp\\
\texttt{HI.20241005.57942.68.fits.gz} \relax& Calibration \relax& ThAr arc lamp\\
\texttt{HI.20241005.57988.70.fits.gz} \relax& Calibration \relax& ThAr arc lamp\\
\bottomrule
\end{longtable}

\LTcapwidth=\columnwidth
\begin{longtable}{lll}
\caption{Data files from 2024 October 21 used in this study and their corresponding observation type and target. All Io observations were taken during eclipse.}\label{tab:files-2024-10-21}
\endfirsthead
\toprule
KOA Unique File Name \relax& Type \relax& Target\\
\midrule
\texttt{HI.20241021.34813.39.fits.gz} \relax& Calibration \relax& None (bias)\\
\texttt{HI.20241021.34857.39.fits.gz} \relax& Calibration \relax& None (bias)\\
\texttt{HI.20241021.34901.62.fits.gz} \relax& Calibration \relax& None (bias)\\
\texttt{HI.20241021.34947.10.fits.gz} \relax& Calibration \relax& None (bias)\\
\texttt{HI.20241021.34991.38.fits.gz} \relax& Calibration \relax& None (bias)\\
\texttt{HI.20241021.35036.26.fits.gz} \relax& Calibration \relax& None (bias)\\
\texttt{HI.20241021.35081.14.fits.gz} \relax& Calibration \relax& None (bias)\\
\texttt{HI.20241021.35125.00.fits.gz} \relax& Calibration \relax& None (bias)\\
\texttt{HI.20241021.35169.88.fits.gz} \relax& Calibration \relax& None (bias)\\
\texttt{HI.20241021.35214.25.fits.gz} \relax& Calibration \relax& None (bias)\\
\midrule
\texttt{HI.20241021.43468.10.fits.gz} \relax& Science \relax& Io\\
\texttt{HI.20241021.43695.50.fits.gz} \relax& Science \relax& Europa\\
\texttt{HI.20241021.44001.56.fits.gz} \relax& Science \relax& Io\\
\texttt{HI.20241021.44234.12.fits.gz} \relax& Science \relax& Europa\\
\texttt{HI.20241021.44769.11.fits.gz} \relax& Science \relax& Io\\
\texttt{HI.20241021.45114.89.fits.gz} \relax& Science \relax& Europa\\
\texttt{HI.20241021.45175.58.fits.gz} \relax& Science \relax& Io\\
\texttt{HI.20241021.45520.86.fits.gz} \relax& Science \relax& Europa\\
\texttt{HI.20241021.46026.27.fits.gz} \relax& Science \relax& Io\\
\texttt{HI.20241021.46377.39.fits.gz} \relax& Science \relax& Europa\\
\texttt{HI.20241021.47393.70.fits.gz} \relax& Science \relax& Jupiter\\
\texttt{HI.20241021.55794.30.fits.gz} \relax& Science \relax& HD 21686\\
\midrule
\texttt{HI.20241021.57041.76.fits.gz} \relax& Calibration \relax& Quartz flat lamp\\
\texttt{HI.20241021.57087.15.fits.gz} \relax& Calibration \relax& Quartz flat lamp\\
\texttt{HI.20241021.57132.54.fits.gz} \relax& Calibration \relax& Quartz flat lamp\\
\texttt{HI.20241021.57177.93.fits.gz} \relax& Calibration \relax& Quartz flat lamp\\
\texttt{HI.20241021.57223.32.fits.gz} \relax& Calibration \relax& Quartz flat lamp\\
\texttt{HI.20241021.57268.71.fits.gz} \relax& Calibration \relax& Quartz flat lamp\\
\texttt{HI.20241021.57314.10.fits.gz} \relax& Calibration \relax& Quartz flat lamp\\
\texttt{HI.20241021.57359.49.fits.gz} \relax& Calibration \relax& Quartz flat lamp\\
\texttt{HI.20241021.57404.88.fits.gz} \relax& Calibration \relax& Quartz flat lamp\\
\texttt{HI.20241021.57449.76.fits.gz} \relax& Calibration \relax& Quartz flat lamp\\
\texttt{HI.20241021.57519.12.fits.gz} \relax& Calibration \relax& ThAr arc lamp\\
\texttt{HI.20241021.57564.51.fits.gz} \relax& Calibration \relax& ThAr arc lamp\\
\texttt{HI.20241021.57609.90.fits.gz} \relax& Calibration \relax& ThAr arc lamp\\
\texttt{HI.20241021.57655.29.fits.gz} \relax& Calibration \relax& ThAr arc lamp\\
\texttt{HI.20241021.57700.68.fits.gz} \relax& Calibration \relax& ThAr arc lamp\\
\texttt{HI.20241021.57746.70.fits.gz} \relax& Calibration \relax& ThAr arc lamp\\
\texttt{HI.20241021.57791.97.fits.gz} \relax& Calibration \relax& ThAr arc lamp\\
\texttt{HI.20241021.57837.39.fits.gz} \relax& Calibration \relax& ThAr arc lamp\\
\texttt{HI.20241021.57883.77.fits.gz} \relax& Calibration \relax& ThAr arc lamp\\
\texttt{HI.20241021.57929.16.fits.gz} \relax& Calibration \relax& ThAr arc lamp\\
\bottomrule
\end{longtable}

\end{document}